\documentclass[12pt]{article}
\usepackage{graphicx}
\usepackage{epsfig}
\usepackage{subfigure}
\usepackage{psfrag}
\usepackage{amssymb}
\usepackage{amsfonts}
\usepackage{amsmath}
\usepackage{bbold}
\allowdisplaybreaks[1]
\newcommand{\be}{\begin{equation}}
\newcommand{\ee}{\end{equation}}
\newcommand{\bea}{\begin{eqnarray}}
\newcommand{\eea}{\end{eqnarray}}
\newcommand{\bfig}{\begin{figure}}
\newcommand{\efig}{\end{figure}}
\newcommand{\bc}{\begin{center}}
\newcommand{\ec}{\end{center}}
\newcommand{\bd}{\begin{displaymath}}
\newcommand{\ed}{\end{displaymath}}

\newcommand{\afb}{A_{\mbox{\tiny{FB}}}}
\newcommand{\as}{\alpha_s}
\newcommand{\ord}{\mathcal{O}}

\newcommand{\sia}{\sigma_A}
\newcommand{\sis}{\sigma_S}

\newcommand{\nn}{\nonumber}

\newcommand{\veg}{v_e^\gamma}
\newcommand{\vqg}{v_Q^\gamma}
\newcommand{\vez}{v_e^Z}
\newcommand{\vqz}{v_Q^Z}
\newcommand{\aez}{a_e^Z}
\newcommand{\aqz}{a_Q^Z}
\renewcommand{\Re}{\mathrm{Re}\;\!}
\renewcommand{\Im}{\mathrm{Im}\;\!}
\newcommand{\aas}{\left(\frac{\alpha_s}{2\pi}\right)}
\newcommand{\hv}{{\hat v}}
\newcommand{\ha}{{\hat a}}
\begin{document}

%
%
%
\begin{titlepage}

\nopagebreak
{\flushright{
        \begin{minipage}{5cm}
        PITHA 06/02 \\
        IFIC/06-09\\
        ZU-TH 08/06 \\
        UCLA/06/TEP/07\\
        \end{minipage}        }
}
\vskip 1.5cm
\begin{center}
{\Large \bf Two-Parton Contribution to the Heavy-Quark\\
\vspace*{2mm}
Forward-Backward Asymmetry in NNLO QCD}
\vskip 1.2cm
{\large  W.~Bernreuther$\rm \, ^{a }$},
{\large  R.~Bonciani$\rm \, ^{b }$},
{\large T.~Gehrmann$\rm \, ^{c }$}, 
{\large R. Heinesch $\rm \, ^{a,\, }$\footnote{present address:
    Framatome ANP GmbH, D-63067 Offenbach, Germany}},  \\[2mm] 
{\large T.~Leineweber$\rm \, ^{a, \, }$\footnote{present address: Framatome
ANP GmbH,
D-91050 Erlangen, Germany}} ,
{\large P. Mastrolia$\rm \, ^{c, \, d}$} 
and {\large E.~Remiddi$\rm \, ^{e}$}
\vskip .6cm
{\it $\rm ^a$ Institut f\"ur Theoretische Physik, RWTH Aachen,
D-52056 Aachen, Germany} 
\vskip .2cm
{\it $\rm ^b$  Departament de F\'{\i}sica Te\`orica, IFIC, CSIC -- Universitat de 
Val\`encia,\\ E-46071 Val\`encia,
Spain} 
\vskip .2cm
{\it $\rm ^c$ Institut f\"ur Theoretische Physik, 
Universit\"at Z\"urich, CH-8057 Z\"urich, Switzerland}
\vskip .2cm
{\it $\rm ^d$ Department of Physics and Astronomy, UCLA, 
Los Angeles, CA 90090-1547, USA}
\vskip .2cm
{\it $\rm ^e$ Dipartimento di Fisica dell'Universit\`a di Bologna, and
INFN, Sezione di Bologna, \\ I-40126 Bologna, Italy} 
\end{center}
\vskip .4cm

\begin{abstract}  
Forward-backward asymmetries, $A_{FB}^Q$,  are important observables for the
determination of the neutral-current couplings of heavy quarks  
in inclusive heavy quark production, $e^+ e^- \to \gamma^*, Z^* \to Q
+X$. In view of the measurement perspectives on  $A_{FB}^Q$ 
at a future linear collider,  precise predictions of $A_{FB}^Q$ are required
for massive quarks. We compute the 
contribution of the $Q \bar Q$ final state to  $A_{FB}^Q$
to order $\as^2$ in the QCD coupling. We
provide general formulae, and we show that this contribution to
$A_{FB}^Q$ is infrared-finite. We evaluate these two-parton
contributions for $b$ and $c$ quarks on and near the $Z$ resonance, and for
$t$ quarks above threshold. Moreover, near the
$t \bar t$ threshold we obtain,  by expanding in the 
heavy-quark velocity $\beta$, an expression for 
$A_{FB}^{t \bar t}$ to order $\as^2$ and NNLL in  $\beta$.
This quantity is equal, to this order in $\beta$,
to the complete forward-backward asymmetry $A_{FB}^t$.
\vskip .3cm
\flushright{
        \begin{minipage}{12.3cm}
\hspace*{18.5mm}  Heavy quarks, Precision calculations, QCD corrections. 
        \end{minipage}        }
\end{abstract}
\vfill
\end{titlepage}    

%
%
%
%
%
\setcounter{footnote}{0}

 \section{Introduction \label{Intro}}

Forward-backward asymmetries  in the production
of fermions in high-energy $e^+ e^-$ collisons are known to be  precision
obervables for the determination  of the respective fermionic
neutral current couplings. Specifically the forward-backward asymmetry
$A_{FB}^b$
of $b$ quarks, which was measured at the $Z$ resonance with an accuracy of 1.7 percent,
led to a determination of the effective weak mixing angle  
$\sin^2 \theta_{W,eff}$ of the Standard Model with a relative precision
of about 1 per mille  \cite{:2004qh,Abbaneo:1998xt} -- notwithstanding the apparent
discrepancy between this measurement and the determination of $\sin^2
\theta_{W,eff}$ with similar precision from the left-right asymmetry
measured by the SLD collaboration. (For a comprehensive overview, see
\cite{:2004qh}.)

At a future linear $e^+ e^-$ collider \cite{tesla}, precision
determinations of electroweak parameters will again involve forward-backward asymmetries.
When such a collider will be operated at the $Z$ peak,  accuracies of
about 0.1 percent may be reached for these observables
\cite{Hawkings:1999ac,Erler:2000jg}. 
Moreover, the top quark asymmetry $A_{FB}^t$ will be experimentally
accessible -- a crucial tool for 
the determination  of the hitherto unexplored neutral
current couplings of this quark.

The theoretical understanding of these observables, in
particular those involving the heavy quarks $Q=t,b,c$, must eventually match these
projected accuracies. The present theoretical description of
$A_{FB}^b$ and $A_{FB}^c$ includes the fully massive next-to-leading
order (NLO) electroweak \cite{Bohm:1989pb,Bardin:1999yd,Freitas:2004mn}
and fully massive NLO QCD \cite{Jersak:1981sp,Arbuzov,Djouadi} corrections. The
next-to-next-to-leading
order (NNLO) QCD corrections, i.e., the contributions of  $\as^2$ to  these
asymmetries,  were calculated so far only in the limit of massless
quarks $Q$ \cite{Altarelli,vanNeerven,seymour}. To be precise, 
 the forward-backward asymmetry of
{\it massless} quarks is not computable in QCD perturbation theory, as was
pointed out in \cite{seymour}.  It is
affected in the limit $m_Q \to 0$ by logarithmic final state
divergences $\sim \ln m_Q$ resulting from the contribution of 
the $Q{\bar Q}Q{\bar Q}$ final state. (These terms are associated with
the non-pertubative $Q$ fragmentation function.) These logarithmically
enhanced terms were taken into account in Ref. \cite{seymour}, which is
the most complete calculation of $A_{FB}^b$
within QCD to date, done both with respect to the
quark and the thrust axis.

In view of the future
perspectives for the $b$- and $t$-quark asymmetries 
at a linear collider,  a computation of  the order $\as^2$ contributions to
$A_{FB}^Q$ for massive quarks $Q$ is clearly desirable. 
The NNLO QCD corrections
involve three classes of contributions: (1) the two-loop corrections
to the decay of a vector boson into a 
heavy quark-antiquark pair; (2) the one-loop corrected matrix elements for 
the decay of a vector boson into a heavy quark-antiquark pair plus a gluon;
(3) the tree level matrix elements for 
the decay of a vector boson into four partons, at least two of which being 
the heavy quark-antiquark pair.

In the limit of massless external quarks $Q$ the $Q \bar Q$
contributions to $A_{FB}^Q$ vanish  up to a non-universal correction
of order $\as^2$ due to quark triangle diagrams \cite{Altarelli}. This
is no longer the case for $m_Q \neq 0$. 
In this paper we determine class (1), i.e.,  the order
$\as^2$ $Q \bar Q$ contributions to $A_{FB}^Q$, for arbitrary quark mass $m_Q$ and
center-of-mass (c.m.) energy $\sqrt{s}$. For this purpose we use
our recent results on the two-loop vector and axial vector vertex
functions for massive quarks \cite{us1,us2,us3}
which determine the amplitude of $e^+e^-\to \gamma^*, Z^* \to Q {\bar
  Q}$ to order $\as^2$ and to lowest order in the electroweak couplings. 
The contributions (1) from the two-parton final state, and (2) plus
(3), i.e., those from the three- and four-parton final states, are separately
infrared-finite, as will be discussed below. The latter can be
obtained along the lines of the calculations
of three-jet production involving heavy quarks
\cite{Bernreuther:1997jn,Rodrigo:1997gy,Nason:1997tz}.  The
parity-violating part of $d\sigma(3\, jet)$ at order $\as^2$, which is
a necessary ingredient here, was computed in \cite{Bernreuther:2000zx}.
However, a full computation of $A_{FB}^{3+4 \, parton}$ has not yet
been done for massive quarks.

The paper is organized as follows. In Section 2 we set up the formulae
for determining $A_{FB}^Q$,  off and at the $Z$ resonance, to order
$\as^2$ in terms of the symmetric and antisymmetric cross sections for
inclusive production of quarks $Q$. In particular we express the $Q \bar
Q$ contribution to $A_{FB}^Q$ by the one- and two-loop heavy-quark vector and
axial vector form factors that determine the amplitude corresponding to
the  $Q \bar Q$ final state. Moreover, we show that these
contributions to the respective
forward-backward asymmetry, which we  denote by
$A_{FB}^{Q\bar Q}$,  are infrared-finite. In Section 3 we present the NNLO
$Q \bar Q$ cross section and the forward-backward asymmetry 
 in the energy region $\alpha_s \ll \beta \ll 1$, where
$\beta$ denotes the heavy-quark velocity.
This result
is applicable, for instance,  to $t$ quarks in the vicinity 
of the $t \bar t$ threshold. 
In Section 4 we
evaluate $A_{FB}^{Q\bar Q}$ for $b$ and $c$ quarks at the $Z$
resonance, and in the case of $t$ quarks for c. m. energies above  the $t\bar t$
threshold till 1 TeV. We conclude in Section 5.

%
%
%
\section{The Forward-Backward Asymmetry to Order $\alpha_s^2$}

In this paper we consider the production of a heavy
quark-antiquark pair in $e^+ e^-$ collisions, 
\begin{align}
e^+ e^- \rightarrow \gamma^*(q),\,Z^*(q) \rightarrow Q\bar{Q} +
X\label{eeQQX} \, ,
\end{align}
where $Q=c,b,t$, to lowest order in the electroweak couplings 
and to second order in
the QCD coupling $\alpha_s$.  To this order the  
following final states  contribute to the cross section of 
inclusive $Q \bar Q$ production, 
Eq.~(\ref{eeQQX}):
the two-parton $Q \bar Q$ state (at Born level, to order $\alpha_s$, 
and to order $\alpha_s^2$), the three-parton state  $Q {\bar Q} g$  (to
order $\alpha_s$ and to order $\alpha_s^2$)
and the  four-parton states $Q{\bar Q}gg$,  $Q{\bar Q} q \bar q$, and
$Q{\bar Q}Q{\bar Q}$ (to order  $\alpha_s^2$).

The forward-backward asymmetry\footnote{We drop the superscript $Q$ in
  the following for ease of notation.} $\afb$ for a heavy quark $Q$ is
commonly defined as the number of quarks $Q$ observed in the forward hemisphere
minus the number of quarks $Q$ in the backward hemisphere, divided by
the total number of observed quarks $Q$. 
The axis that defines the forward direction must be 
infrared- and collinear-safe in order that   $\afb$ is computable
in perturbation theory. Common choices are the direction of flight of
$Q$ or the thrust axis direction. The forward-backward asymmetry can
also be expressed in terms of the cross section for inclusive
production of quarks $Q$.  
We have:
\begin{align}
\afb = \frac{\sia}{\sis}\label{defafb},
\end{align}
with the antisymmetric and symmetric cross sections $\sigma_A$ and
$\sigma_S = \sigma$ defined by
\begin{align}
\sia = & \int_0^1\frac{d\sigma}{d\cos\vartheta} d\cos\vartheta - \int_{-1}^0\frac{d\sigma}{d\cos\vartheta} d\cos\vartheta\label{defsia},\\
\sis = & \int_{-1}^1\frac{d\sigma}{d\cos\vartheta} d\cos\vartheta \, .
\label{defsis}
\end{align}
Here $\vartheta$ is the angle between the incoming electron and the 
direction defining the forward hemisphere (in the $e^+ e^-$ center-of-mass
frame). When choosing the momentum direction of $Q$ or the
thrust axis, the $Q \bar Q$ contribution to $\afb$, which we compute in
this paper, is of course the same. In the following  $\vartheta = \angle(e^-,Q)$.

In analogy to its experimental measurement, $\afb$ may be computed by
determining the contributions from the final-state jets which, to order
$\as^2$, are those of
the two-, three-, and four-jet states.  These contributions
are separately infrared-finite, and $\afb$ would not depend  on the jet
clustering algorithm employed when no phase space cuts are applied.
Such a calculation would require a jet calculus for massive
quarks
at NNLO in $\as$  which is, however, not available. (For massless
quarks a NNLO subtraction method was recently developed
\cite{Gehrmann-DeRidder:2005cm},  cf. also references therein.)
Yet,  $\afb$ belongs to the class of observables that can be computed
at the level of unresolved partons. The two-parton and the three- plus
four-parton contributions to the second-order forward-backward
asymmetry are separately infrared (IR) finite,
cf. \cite{Altarelli,seymour}
 and Section 2.2 below. This basic result will be exploited in the following.

To order $\alpha_s^2$ the  symmetric and antisymmetric  cross sections
receive the following contributions from unresolved partons:
\begin{align}\label{sasbeitraege}
\sigma_{A,S} = \sigma_{A,S}^{(2,0)} +\sigma_{A,S}^{(2,1)} + \sigma_{A,S}^{(2,2)} + \sigma_{A,S}^{(3,1)} + \sigma_{A,S}^{(3,2
)} + \sigma_{A,S}^{(4,2)} + \ord(\as^3) \, ,
\end{align}
where the first number in  the superscripts $(i,j)$  denotes the
number of partons in the respective 
final state and the second one the order  of
$\alpha_s$. Inserting (\ref{sasbeitraege}) into (\ref{defafb}) we get
for $\afb$ to second order in $\as$:
\begin{align}\label{afbgesamt}
\afb(\as^2) = \frac{\sigma_{A}^{(2,0)} +\sigma_{A}^{(2,1)} + \sigma_{A}^{(2,2)} + \sigma_{A}^{(3,1)} + \sigma_{A}^{(3,2)} +
\sigma_{A}^{(4,2)}}{\sigma_{S}^{(2,0)} +\sigma_{S}^{(2,1)} + \sigma_{S}^{(2,2)} + \sigma_{S}^{(3,1)} + \sigma_{S}^{(3,2)} +
\sigma_{S}^{(4,2)}} \, .
\end{align}
Taylor expansion of~\eqref{afbgesamt} with respect to  $\as$ leads to 
\begin{align}\label{afbent}
\afb(\as^2) =&
A_{\mbox{\tiny{FB}},0}\;\left[1\;+\;A_{1} \;+\;
A_{2}\right] \, ,
\end{align}
where $A_{\mbox{\tiny{FB}},0}$ is
the forward-backward asymmetry at Born level, and  $A_{1}$ and $A_{2}$
are the $\ord(\as)$ and  $\ord(\as^2)$
contributions  normalized to $A_{\mbox{\tiny{FB}},0}$. 
\begin{align}
A_{\mbox{\tiny{FB}},0} =& \frac{\sia^{(2,0)}}{\sis^{(2,0)}} \, ,\label{afb0}\\
A_{1} =& \frac{\sia^{(2,1)}}{\sia^{(2,0)}} \;-\;
\frac{\sis^{(2,1)}}{\sis^{(2,0)}} \;+\; \frac{\sia^{(3,1)}}{\sia^{(2,0)}} \;-\;
\frac{\sis^{(3,1)}}{\sis^{(2,0)}} \, ,\label{afb1}\\
A_{2} =& \frac{\sia^{(2,2)}}{\sia^{(2,0)}} \;-\;
\frac{\sis^{(2,2)}}{\sis^{(2,0)}}
+ \frac{\sia^{(3,2)}}{\sia^{(2,0)}} \;-\;
\frac{\sis^{(3,2)}}{\sis^{(2,0)}} \;+\;
\frac{\sia^{(4,2)}}{\sia^{(2,0)}} \;-\;
\frac{\sis^{(4,2)}}{\sis^{(2,0)}}\nonumber\\
-&\frac{\sis^{(2,1)}+\sis^{(3,1)}}{\sis^{(2,0)}} \;\Bigg[\;\frac{\sia^{(2,1)}}{\sia^{(2,0)}} \;-\;
\frac{\sis^{(2,1)}}{\sis^{(2,0)}} \;+\;
\frac{\sia^{(3,1)}}{\sia^{(2,0)}} \;-\;
\frac{\sis^{(3,1)}}{\sis^{(2,0)}}\;\Bigg] \, .\label{afb2}
\end{align}
\subsection{The $Q \bar Q$ Contribution}

It is convenient to rewrite  Eq. (\ref{afbent}) as follows:
\be
\afb(\as^2) = 
A^{(2p)}_{\mbox{\tiny{FB}}} \;+\;
A^{(3p)}_{\mbox{\tiny{FB}}}  \;+\; A^{(4p)}_{\mbox{\tiny{FB}}}  \, ,
\label{afp234}
\ee
where the superscript $(np)$ labels the number of partons $n$ in the final
state. Collecting the two-parton term from
Eqs.~\eqref{afb0},~\eqref{afb1},  
and~\eqref{afb2} we get:
\be
A^{(2p)}_{\mbox{\tiny{FB}}} = A_{\mbox{\tiny{FB}},0}\; 
\left[1 \; + \; A^{(2p)}_1 +
  A^{(2p)}_2 \right ] \, ,
\label{afb2pp}
\ee
with
\begin{align}
A_{1}^{(2p)} =& \frac{\sia^{(2,1)}}{\sia^{(2,0)}}\;-\;\frac{\sis^{(2,1)}}{\sis^{(2,0)}},\label{afb12p}\\
A_{2}^{(2p)} =& A_{2,2} - A_{2,1}, \label{afb22p}
\end{align}
where
\begin{align}
A_{2,2} =& \frac{\sia^{(2,2)}}{\sia^{(2,0)}} \;-\;
\frac{\sis^{(2,2)}}{\sis^{(2,0)}}, \label{afb2loop} \\
A_{2,1} = & \frac{\sis^{(2,1)}}{\sis^{(2,0)}} \; A_{1}^{(2p)}\, .
\label{afb21loop}
\end{align}
The remaining terms in Eqs.~\eqref{afb1},~\eqref{afb2} contribute to 
$A^{(3p)}$ and $A^{(4p)}$.

As already stated above,
both the two-parton  and the three- plus four-parton contributions $A^{(2p)}$
and $A^{(3p)}+ A^{(4p)}$, respectively, are infrared (IR)
finite, i.e., free of soft and collinear singularites. We shall
show this  explicitly for $A^{(2p)}$ at the end of this section.

Next we express $A^{(2p)}_{\mbox{\tiny{FB}}}$ in terms of the $VQ \bar
Q$  vertex form factors $(V=\gamma,Z)$ which determine the  amplitude 
of the reaction  $e^+ e^- \rightarrow \gamma^*,\,Z^* \rightarrow Q\bar{Q}$  
to lowest order in the electroweak couplings and to any order in the
QCD coupling; see Fig.~1.
\begin{figure}[ht]
\begin{center}
\epsfig{file=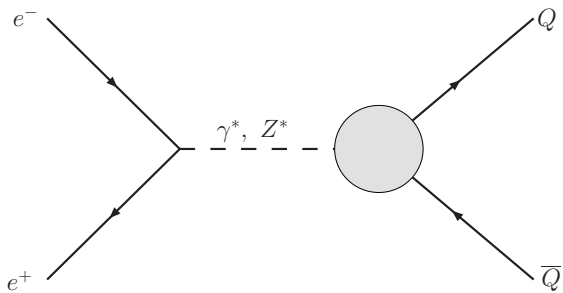, width=8cm, height=4cm}
\end{center}
\caption{\label{skanal} The amplitude $e^+ e^- \to Q {\bar Q}$  in QCD.}
\end{figure}
In this case the  $VQ \bar Q$ vertex $\Gamma^{\mu,V}_Q$  depends,
for on-shell external quarks, on  four form factors:
\bea
i\, \Gamma^{\mu,V}_Q  &=&   v_Q^V \left (F_{1}(s) \gamma^\mu +
 \frac{i}{2m_Q}F_{2}(s)\sigma^{\mu \nu} q_\nu \right)
\nn \\ && +  a_Q^V \left(G_{1}(s) \gamma^\mu \gamma_5 +
  \frac{1}{2m_Q}G_{2}(s) \gamma_5\, q^\mu \right) \, ,\label{decomp}
\eea
where 
$s=q^2$, $\sigma^{\mu \nu} = \frac{i}{2}\left[\gamma^\mu,\gamma^\nu
\right]$,  $m_Q$ denotes the on-shell mass of $Q$, and, for $f=e, Q$,
\begin{align}
v_f^Z =& \,\frac{e}{2 s_W
  c_W}\,\left(T^3_f- 2s^2_W \, e_f\right) \, ,  \nn\\
a_f^Z =& \,\frac{e}{2 s_W c_W} \,\left(-T^3_f\right) \, , \nn\\
v_f^\gamma =& \,e \,e_f\, , \quad a_f^\gamma = 0 \, . 
\label{vakoppl}
\end{align}
Here $e_f$ and $T^3_f$ denote the charge of $f$ in units of the
positron charge $e$ and
its weak isospin, respectively, and $s_W\,(c_W)$ are the sine
(cosine) of  the weak mixing angle 
$\theta_W$. The
functions $F_i$ and $G_i$ denote renormalized form factors; the 
renormalization scheme will be specified below. Instead of
using the Dirac form factor $F_1$ we express in the following 
the (anti)symmetric cross
section in terms of 
\begin{align}
\tilde{F}_1(s) = F_1(s) + F_2(s) \, . 
\end{align}
Neclecting the electron
 mass we find for the two-parton contributions to $\sigma_{A,S}$:
\begin{align}
\sia^{(2p)} = &\frac{N_c}{8\pi}\,\frac{s}{D_Z} \, \beta^2 \, \aez \aqz \Bigg\{\left[\vez\vqz + \frac{1}{2}\left(1-\frac{m_Z^
2}{s}\right)\,\veg\vqg\right]\left(\tilde{F}_1^* G_1 + \tilde{F}_1 G_1^*\right) \nn\\
& + i \frac{m_Z \Gamma_Z}{2 s}\vqg \veg \left(\tilde{F}_1 G_1^* -
  \tilde{F}_1^* G_1\right) \Bigg\}\, ,\label{siaff}
\end{align}
\begin{align}
\sis^{(2p)} = &
\frac{N_c}{24\pi} \,\frac{1}{s}\,\beta\,\left(\veg\vqg \right)^2 \; \mathcal{W} + \frac{N_c}{12\pi} \,\frac{s}{D_Z} \,\beta
\,\left(1-\frac{m_Z^2}{s}\right)\,\veg\vqg\vez\vqz \; \mathcal{W} \nn\\
+ & \frac{N_c}{24\pi} \,\frac{s}{D_Z}\, \beta\,
\left[\left(\aez\right)^2 + \left(\vez\right)^2\right]\;
\Big[\left(\vqz\right)^2 \; \mathcal{W} \; +
2\beta^2\left(\aqz\right)^2 \left(G_1 G_1^*\right)\Big]\, , \label{sisff}
\end{align}
where
\begin{align}
\mathcal{W} = (3-\beta^2)\,\left(\tilde{F}_1
  \tilde{F}_1^*\right)+\beta^2 \left(\tilde{F}_1 F_2^* + \tilde{F}_1^*
  F_2\right) + \frac{\beta^4}{1-\beta^2}\left(F_2 F_2^*\right)\, ,\nn
\end{align}
$D_Z = [(s-m_Z^2)^2 + m_Z^2 \Gamma_Z^2]$ with $m_Z,$ $\Gamma_Z$ being
the mass and width of the  $Z$ boson,
 $\beta = \sqrt{1-\frac{4m_Q^2}{s}}$ the heavy quark velocity, and
 $N_c=3$ the number of colors. Because we have put $m_e=0$ 
the form factor $G_2$ does not contribute to
Eqs.~(\ref{siaff}),~(\ref{sisff}). 
The last term in~\eqref{siaff}, which contains  $\Gamma_Z$,  is of higher order in the
electroweak couplings as compared with the first term.  We will neglect 
that term in the following.\\
Expanding the form factors in 
Eqs.~(\ref{siaff}),~(\ref{sisff}) in powers of $\as$:
\begin{alignat}{3}
\tilde{F}_1 = & 1 \;+\; &\aas \;\tilde{F}_1^{(1\ell)} &+ \aas^2 \;\tilde{F}_1^{(2\ell)} & \;+\; \ord\left(\as^3\right), \label{f1ent}\\
F_2 = && \aas \;F_{2}^{(1\ell)} &+ \aas^2 \;F_{2}^{(2\ell)} &\;+\; \ord\left(\as^3\right), \label{f2ent}\\
G_1 = & 1 \;+\; &\aas \;G_{1}^{(1\ell)} &+ \aas^2 \;G_{1}^{(2\ell)} &\;+\; \ord\left(\as^3\right), \label{g1ent}
\end{alignat}
leads to the expansions of the two-parton contributions as exhibited
in~\eqref{sasbeitraege}, i.e., 
\begin{align}
\sigma_{A,S}^{(2p)} = & \, \sigma_{A,S}^{(2,0)} +\sigma_{A,S}^{(2,1)} +
\sigma_{A,S}^{(2,2)}. \label{sas22p}
\end{align}
The contributions to $\sigma_{S}^{(2p)}$  can be further decomposed as follows:
\begin{alignat}{3}
\sis^{(2,0)} = & \;\sis^{(2,0,\gamma)} &&+ \sis^{(2,0,Z)} &&+\sis^{(2,0,\gamma Z)},\\
\sis^{(2,1)} = & \;\sis^{(2,0,\gamma)} \sis^{(2,1,\gamma)} &&+ \sis^{(2,0,Z)} \sis^{(2,1,Z)} &&+ \sis^{(2,0,\gamma Z)} \sis^
{(2,1,\gamma Z)},\\
\sis^{(2,2)} = & \;\sis^{(2,0,\gamma)} \sis^{(2,2,\gamma)} &&+ \sis^{(2,0,Z)} \sis^{(2,2,Z)} &&+ \sis^{(2,0,\gamma Z)} \sis^{(2,2,\gamma Z)},
\label{s2loops}
\end{alignat}
where the superscripts  $\gamma$, $Z$, and $\gamma Z$
label  the pure photon and $Z$ contributions, and the
$\gamma-Z$ interference terms, respectively. These terms are given by

\begin{align}
\sis^{(2,0,\gamma)} = & \frac{N_c}{24\pi} \,\frac{1}{s}\,\beta\,\left(\veg\vqg \right)^2 (3-\beta^2),\label{ss20g}\\
\sis^{(2,1,\gamma)} = & \aas \Big\{2 \Re \tilde{F}_1^{(1\ell)} + \frac{2\beta^2}{3-\beta^2} \Re F_2^{(1\ell)}\Big\},\label{ss21gff}\\
\sis^{(2,2,\gamma)} = & \aas^2 \Big\{\frac{2\beta^2}{3-\beta^2}\Big[ \Re F_2^{(2\ell)} + \Re \tilde{F}_1^{(1\ell)} \Re F_2^{(1\ell)} + \pi^2\, \Im \tilde{F}_1^{(1\ell)}\Im F_2^{(1\ell)}\Big]\nn\\
& + \frac{\beta^4}{(3-\beta^2)(1-\beta^2)}\Big[ (\Re F_2^{(1\ell)})^2 + \pi^2\, (\Im F_2^{(1\ell)})^2 \Big]\nn\\
& + (\Re \tilde{F}_1^{(1\ell)})^2 + \pi^2\, (\Im \tilde{F}_1^{(1\ell)})^2 + 2 \Re \tilde{F}_1^{(2\ell)}\Big\},\label{ss22gff}\\
\sis^{(2,0,Z)} = & \frac{N_c}{24\pi} \,\frac{s}{D_Z}\, \beta\,
\left[\left(\aez\right)^2 + \left(\vez\right)^2\right]\;\left[2\left(\aqz\right)^2 \beta^2 + \left(\vqz\right)^2 (3-\beta^2)\right],\label{ss20z}\\
\sis^{(2,1,Z)} = & \frac{ \left(\vqz\right)^2 (3-\beta^2)\; \sis^{(2,1,\gamma)} + 4 \aas \left(\aqz\right)^2 \beta^2 \Re G_1^{(1\ell)} }{2\left(\aqz\right)^2 \beta^2 + (3-\beta^2)\left(\vqz\right)^2},\label{ss21zff}\\
\sis^{(2,2,Z)} = & \frac{1}{2\left(\aqz\right)^2 \beta^2 + (3-\beta^2)\left(\vqz\right)^2} \Big\{ \left(\vqz\right)^2 (3-\beta^2)\; \sis^{(2,2,\gamma)}\nn\\
& + 4 \aas^2 \left(\aqz\right)^2 \beta^2 \Re G_1^{(2\ell)} + 2 \aas^2
\left(\aqz\right)^2 \beta^2\Big[(\Re G_1^{(1\ell)})^2 + \pi^2\,(\Im
G_1^{(1\ell)})^2 \Big] \Big\},\label{ss22zff}\\
\sis^{(2,0,\gamma Z)} = & \frac{N_c}{12\pi} \,\frac{s}{D_Z} \,\beta
\,\left(1-\frac{m_Z^2}{s}\right)\,\veg\vqg\vez\vqz \;(3-\beta^2),\label{ss20gz}\\
\sis^{(2,1,\gamma Z)} = & \sis^{(2,1,\gamma)},\label{ss21gzff}\\
\sis^{(2,2,\gamma Z)} = & \sis^{(2,2,\gamma)}, \label{ss22gzff}
\end{align}
where  the convention $F_a = \Re F_a + i\pi \Im F_a$,
$G_a = \Re G_a + i\pi \Im G_a$ 
 $(a=1,2)$ is used; i.e., a factor $\pi$ is taken out of the imaginary part. 
The antisymmetric cross section is given by:
\begin{align}
\sia^{(2,0)} = &\; \frac{N_c}{4\pi}\,\frac{s}{D_Z} \, \beta^2 \, \aez \aqz \left[\vez\vqz + \frac{1}{2}\left(1-\frac{m_Z^2}{
s}\right)\,\veg\vqg\right], \label{sa20ff}\\
\sia^{(2,1)} = &\; \sia^{(2,0)}\; \aas \; \left[\Re \tilde{F}_1^{(1\ell)} + \Re G_1^{(1\ell)}\right],\label{sa21ff}\\
\sia^{(2,2)} = &\; \sia^{(2,0)} \; \aas^2 \; \left[\Re \tilde{F}_1^{(2\ell)} + \Re G_1^{(2\ell)} + \Re \tilde{F}_1^{(1\ell)}\Re G_1^{(1\ell)}
 + \pi^2 \Im \tilde{F}_1^{(1\ell)} \Im G_1^{(1\ell)} \right].\label{sa22ff}
\end{align}
With these formuale  the $Q \bar Q$
contribution~\eqref{afb2pp} to  the forward-backward asymmetry is 
expressed in terms of the one- and two-loop 
form factors Eqs.~\eqref{f1ent}~-~\eqref{g1ent}.

The second order term in the expansion~\eqref{g1ent}
of the axial vector form factor, $G_1^{(2\ell)}$, receives so-called
type A and type B
 contributions. Type A contributions are those where the $Z$ boson
 couples directly to the external quark $Q$~\cite{us2},
 while the triangle diagram
contributions, summed over the  quark isodoublets of
the three generations, are called type B~\cite{us3}.
(In the terminology of~\cite{seymour} these correspond to universal and 
non-universal corrections, respectively.)
Among Eqs.~\eqref{ss20g}  -~\eqref{sa22ff} only $\sis^{(2,2,Z)}$ and  $\sia^{(2,2)}$ depend on 
$G_1^{(2\ell)}$. With $G_1^{(2\ell)}= G_1^{(2\ell,A)}+
G_1^{(2\ell,B)}$
we separate  in these terms the type A and B contributions: 
\begin{align}
\sis^{(2,2,Z)} = \sis^{(2,2,Z,A)} + \sis^{(2,2,Z,B)}
\end{align}
with
\begin{align}
\sis^{(2,2,Z,A)}  = \; \sis^{(2,2,Z)}(G_1^{(2\ell)} \rightarrow G_1^{(2\ell,A)}),\\
\sis^{(2,2,Z,B)}  = \; \aas^2\, \frac{\;4\,\left(\aqz\right)^2\,\beta^2\,
  \Re\,G_1^{(2\ell,B)}}{2\left(\aqz\right)^2 \beta^2 +
  (3-\beta^2)\left(\vqz\right)^2} \; , \label{ss22zb}
\end{align}
and
\begin{align}
\sia^{(2,2)} = \sia^{(2,2,A)} + \sia^{(2,2,B)}
\end{align}
with
\begin{align}
\sia^{(2,2,A)}  = \; \sia^{(2,2)}(G_1^{(2\ell)} \rightarrow G_1^{(2\ell,A)}),\\
\sia^{(2,2,B)}  = \; \aas^2\, \sia^{(2,0)}\;\Re\,G_1^{(2\ell,B)}. \label{sa22zb}
\end{align}
Thus (\ref{afb22p}) can be written: 
\be
A^{(2p)}_{2} = A^{(2p,A)}_{2} +
A^{(2p,B)}_{2} \, ,
\label{af2psep}
\ee
where $A^{(2p,A)}_{2}$ is obtained from (\ref{afb22p})
by the substitution  $G_1^{(2\ell)} \rightarrow G_1^{(2\ell,A)}$
in~\eqref{afb22p}, respectively in ~\eqref{afb2loop}, and 
\begin{align}
A_{2}^{(2p,B)} =& \frac{\sia^{(2,2,B)}}{\sia^{(2,0)}}
\;-\; \frac{\sis^{(2,0,Z)}\,\sis^{(2,2,Z,B)}}{\sis^{(2,0)}} \; .\label{afb22zb}
\end{align}
If one evaluates, in the case of $b \bar b$ and $c \bar c$ final
states, Eq.~(\ref{afb22zb}) exactly at the $Z$ resonance and neglects
the contributions from photon exchange, then:
\begin{align}
A_{2}^{(2p,B)}(s=m_Z^2) =\;  \aas^2 \; \Re G_1^{(2\ell,B)} \;\left[\frac{(3-\beta^2)\left(\vqz\right)^2- 2\left(\aqz\right)^2 \beta^2}{(3-\beta^2)\left(\vqz\right)^2 + 2\left(\aqz\right)^2 \beta^2 }\right]\label{afb22zb1}.
\end{align} 
The type B contribution~(\ref{afb22zb})  to the forward-backward
asymmetry is  ultraviolet- and infrared-finite~\cite{us3}.

\subsection{Infrared-Finiteness of $A_{\mbox{\tiny{FB}}}^{(2p)}$}

The renormalized vector and axial vector form factors~\eqref{decomp}  were computed in \cite{us1,us2,us3}
to order $\as^2$ within QCD with $N_f$ massless and one massive
quark $Q$, in a renormalization scheme, which is appropriate
for the case at hand: the wavefunction and the mass of $Q$ are
defined in the on-shell scheme while  $\as$ is defined in
the ${\overline{\rm MS}}$ scheme. The renormalized
form factors still contain IR divergences which are
regulated by dimensional regularization in $D=4-2\epsilon$ dimensions.
At one loop, $F_1^{(1\ell)}$
and $G_1^{(1\ell)}$ contain $1/\epsilon$ poles due to soft virtual
gluons. At two loops, the dominant IR singularities of  $F_1^{(2\ell)}$
and $G_1^{(2\ell, A)}$ are of order $1/\epsilon^2$  due to soft and
collinear massless partons, while $F_2^{(2\ell)}$ has
only $1/\epsilon$ poles. However, when inserting the 
renormalized form factors into the formulae for the order $\as$ and
$\as^2$ contributions to  $A_{\mbox{\tiny{FB}}}^{(2p)}$ the IR
singularities cancel and we obtain a finite result. This is well-known
for $A_{1}^{(2p)}$, c.f. \cite{Djouadi}. 

This cancellation of infrared poles can be seen in a straightforward 
manner if one takes into account the universal structure of the 
infrared poles in the form factors, which can be expressed in terms of 
one-loop and two-loop infrared singularity operators for a massive 
quark-antiquark pair, ${\boldmath I}^{(1)}_{Q\bar Q}(s,\mu=m_Q,\epsilon)$ and
${\boldmath I}^{(2)}_{Q\bar Q}(s,\mu= m_Q,\epsilon)$, where $\mu$ is
the renormalization scale,  as: 
 \begin{eqnarray}
{\cal P}oles \; \left( F_1^{(1\ell)} \right) &=& {\boldmath I}^{(1)}_{Q\bar Q}
F_1^{(0 \ell)} \;, 
\nonumber \\
{\cal P}oles \; \left( F_2^{(1\ell)}\right) &=& 0\;, \nonumber \\
{\cal P}oles \; \left( G_1^{(1\ell)}\right) &=& {\boldmath I}^{(1)}_{Q\bar Q} G_1^{(0
  \ell)}\; ,\\
{\cal P}oles \; \left(F_1^{(2\ell)}\right)  &=& {\boldmath I}^{(2)}_{Q\bar Q} F_1^{(0\ell)} 
+{\boldmath I}^{(1)}_{Q\bar Q} F_1^{(1\ell)} \; ,\nonumber \\
{\cal P}oles \; \left(F_2^{(2\ell)}\right) &=& {\boldmath I}^{(1)}_{Q\bar Q} F_2^{(1\ell)} \;, 
\nonumber \\
{\cal P}oles \; \left(G_1^{(2\ell)}\right)  &=& {\boldmath I}^{(2)}_{Q\bar Q} G_1^{(0\ell)}\; +{\boldmath I}^{(1)}_{Q\bar Q} G_1^{(1\ell)} \; . 
\end{eqnarray}
This factorization of the infrared poles is a well-known feature of 
massless multi-loop amplitudes, where it can be derived from 
exponentiation~\cite{Catani:1998bh,Sterman:2002qn}. 
In massive QED, the same behaviour 
was observed long ago~\cite{Yennie:1961ad}. To the best of our knowledge, 
infrared factorization in massive QCD was established up to now only at 
the one-loop level~\cite{Catani:2000ef}; the above pole structure of the 
form factors suggests that it holds at higher orders as well. Expressions 
for ${\boldmath I}^{(1,2)}_{Q\bar Q}$ can be read off from the explicit 
pole structure of the form 
factors~\cite{us1,us2}. 
Exploiting that 
\begin{displaymath}
F_1^{(0\ell)} = G_1^{(0\ell)} = 1\;,
\end{displaymath}
one finds immediately that the two-parton contribution  (\ref{afb22p})
to the 
forward-backward asymmetry is infrared finite. However,
it should be  kept in mind that the two terms in (\ref{afb22p})  are not separately finite. 

Let us illustrate how this
cancellation occurs in $A_{2}^{(2p)}$ by considering this expression at
the $Z$ resonance, i.e., by taking into account only $Z$ exchange 
contributions and neglecting the photon and 
$\gamma-Z$ interference terms. When inserting the
form factors into Eq.~\eqref{afb2loop}, all the leading $1/\epsilon^2$
singularities cancel, but a subleading divergence remains:
\begin{eqnarray}
A_{2,2}  &=& -\, \left(\frac{\as}{2\pi}\right)^2 \; C_F^2 \; \frac{1}{\epsilon}
\; 4 y \; \frac{ 2 ({a_Q^Z})^2 (1-y)^2 - 3 ({v_Q^Z})^2
  (1+y)^2}{[({a_Q^Z})^2 (1-y)^2 +
  ({v_Q^Z})^2 (1+y)^2 + 2 ({v_Q^Z})^2 y ]\;(1-y^2)^2} \nonumber\\
&& \times \;[(1+y^2)\;  \ln^2(y)\; + \; (1-y^2) \;  \ln(y)]
 \; + \; A_{2,2}^{finite} \, 
\label{twolsing}
\end{eqnarray}
where $y =  ({1-\beta})/({1+\beta})$  and $C_F=(N_c^2-1)/(2N_c)$.
This subleading singularity proportional to $C_F^2$ results from  the
real parts of the one-loop form factors, $F_1^{(1\ell)}$
and $G_1^{(1\ell)}$, and from the contribution of a set of
abelian-type two-loop diagrams (gluon ladder diagram and gluon
vertex diagrams with quark self-energy insertions) to the real parts
of the two-loop
vector and type A axial vector form factors. The
singularity~\eqref{twolsing}
is removed by the second term in~\eqref{afb22p}. While
\be
A_{1}^{(2p)}
=  \frac{\as}{2\pi} \; C_F \; \ln(y) \; 2 y \; \frac{ 2 ({a_Q^Z})^2
  (1-y)^2 - 
3 ({v_Q^Z})^2 (1+y)^2}{[({a_Q^Z})^2 (1-y)^2 +
  ({v_Q^Z})^2 (1+y)^2 + 2 ({v_Q^Z})^2 y ]\;(1-y^2)} 
\ee
is finite, the first term in~\eqref{afb21loop} contains a singularity,
\be
\frac{\sigma_S^{(2,1)}}{\sigma_S^{(2,0)}}  = 
- \, \frac{\as}{2\pi}\; C_F \; \frac{1}{\epsilon}\:
\frac{2(1-y^2 +(1+y)^2 \ln(y))}{1-y^2} \, + \, {\rm finite}\; {\rm terms} \, ,
\ee
and thus the singular part of $A_{2,1}$ cancels 
the singularity in~\eqref{twolsing}. This cancellation pattern remains
also away from the $Z$ resonance.
%

\section{Cross Section and  $A_{\mbox{\tiny{FB}}}$  near Threshold}

In this section we analyze the cross section and  
$A_{\mbox{\tiny{FB}}}^{(2p)}$ for $e^+ e^- \to \gamma^*, Z^* \to Q
{\bar Q}$ near the production threshold. Close to threshold, where the
quark velocity $\beta$ is small, the fixed order perturbative
expansion in $\alpha_s$ of these and other quantities breaks down due
to Coulomb singularities, which must be resummed. In addition,
large logarithms in $\beta$ appear, which may be summed using the
renormalization group  \cite{Beneke:1999qg,Hoang:2001mm}
applied in the framework of 
nonrelativistic effective field theory methods. However, in the energy region where
$\alpha_s \ll \beta \ll 1$, threshold expansions of observables, i.e.,
expansions in $\beta$ to fixed order in  $\alpha_s$  are  expected
to yield reliable results. In this region we derive compact formulae
for the symmetric and antisymmetric $Q \bar Q$ cross sections at NNLO.
These expressions should be useful, especially in  the case of
top-quark pair production, for comparison of continuum results
with results obtained directly at threshold. 

\subsection{The $Q \bar Q$ cross section}
\begin{figure}[ht]
\begin{center}
\epsfig{file=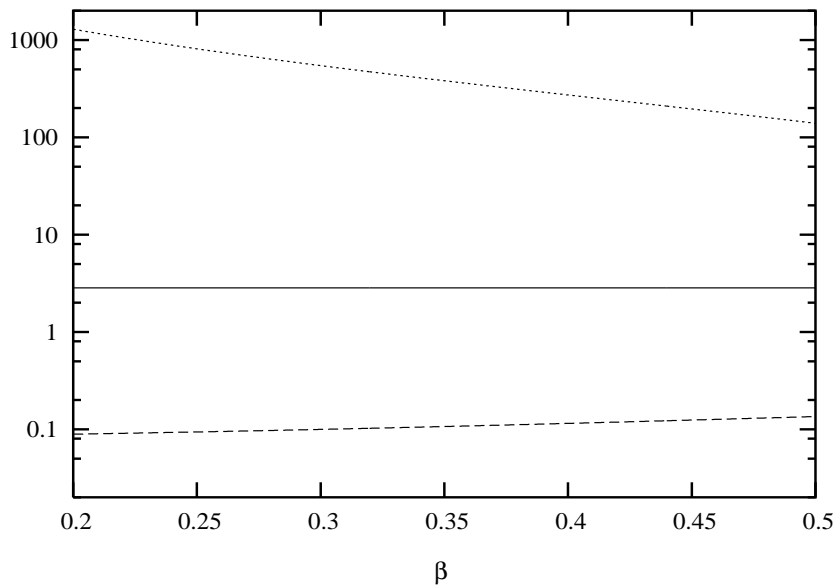, width=12cm, height=8cm}
\end{center}
\caption{\label{thr} The terms $\Delta^{(2,Ve)}$ (dotted line), $\Delta^{(0,Ax)}$ (dashed
line) and $\Delta^{(2,Ax)}$ (solid line), expanded to 
order $\beta^0$, as functions of $\beta$, for the  case of $t \bar t$
production, using $m_t$=172.7 GeV. $\beta=0.2$ ($\beta=0.5$)
corresponds to a  center-of-mass energy of about 357 GeV (404 GeV).}
\end{figure}

We start with Eqs. (\ref{sas22p}) - (\ref{s2loops}) where the $Q \bar
Q$ cross section at NNLO, $\sigma_{NNLO}=\sigma_S^{(2p)}$, is expressed
in terms of the $\gamma,$ $Z$, and $\gamma-Z$ exchange contributions.
Using these formulae we may write
\bea
\sigma_{NNLO} & = &
 \sigma_{S}^{(2,0,\gamma)} \Biggl\{ 1 + \frac{\sigma_{S}^{(2,0,Z)} +
\sigma_{S}^{(2,0,\gamma Z)}}{\sigma_{S}^{(2,0,\gamma)}} 
+ \sigma_{S}^{(2,1,\gamma)} \Biggl[ 1 +
\frac{\sigma_{S}^{(2,0,Z)} \sigma_{S}^{(2,1,Z)} 
+ \sigma_{S}^{(2,0,\gamma Z)} \sigma_{S}^{(2,1,\gamma Z)}}{
\sigma_{S}^{(2,0,\gamma)} \sigma_{S}^{(2,1,\gamma)}} \Biggr] \nn\\
& & \hspace{1.3cm} + \sigma_{S}^{(2,2,\gamma)} \Biggl[ 1
+ \frac{\sigma_{S}^{(2,0,Z)} \sigma_{S}^{(2,2,Z)} 
+ \sigma_{S}^{(2,0,\gamma Z)} \sigma_{S}^{(2,2,\gamma Z)}}{
\sigma_{S}^{(2,0,\gamma)} \sigma_{S}^{(2,2,\gamma)}} \Biggr] \Biggl\} \,.
\eea
Denoting
\be
\Delta^{(0,Ax)} = \frac{\sigma_{S}^{(2,0,Z)} +
\sigma_{S}^{(2,0,\gamma Z)}}{\sigma_{S}^{(2,0,\gamma)}} 
\ee
and recalling Eqs.  (\ref{ss21gzff}), (\ref{ss22gzff})
we get
\bea
\sigma_{NNLO} & = &
\sigma_{S}^{(2,0,\gamma)} \Biggl\{ 1 + \Delta^{(0,Ax)}
+ \sigma_{S}^{(2,1,\gamma)} \Bigl[ 1 +
\Delta^{(0,Ax)} \Bigr] + 
\frac{\sigma_{S}^{(2,0,Z)}}{\sigma_{S}^{(2,0,\gamma)}} \Bigl( 
\sigma_{S}^{(2,1,Z)} - \sigma_{S}^{(2,1,\gamma)} \Bigr) \nn\\
& & \hspace{1.3cm} + \sigma_{S}^{(2,2,\gamma)} \Bigl[ 1 +
\Delta^{(0,Ax)} \Bigr] + 
\frac{\sigma_{S}^{(2,0,Z)}}{\sigma_{S}^{(2,0,\gamma)}} \Bigl(  
\sigma_{S}^{(2,2,Z)} - \sigma_{S}^{(2,2,\gamma)} \Bigr) \Biggl\} \, .
\label{final}
\eea
Putting
\bea
\frac{\sigma_{S}^{(2,0,Z)}}{\sigma_{S}^{(2,0,\gamma)}} \Bigl( 
\sigma_{S}^{(2,1,Z)} - \sigma_{S}^{(2,1,\gamma)} \Bigr) 
&=& C_{F} \left( \frac{\alpha_{s}}{2 \pi} \right) {\Delta}^{(1,Ax)} \, ,
\label{DE1} \\
\frac{\sigma_{S}^{(2,0,Z)}}{\sigma_{S}^{(2,0,\gamma)}} \Bigl(  
\sigma_{S}^{(2,2,Z)} - \sigma_{S}^{(2,2,\gamma)} \Bigr) 
&=& C_{F} \left( \frac{\alpha_{s}}{2 \pi} \right)^2 \Delta^{(2,Ax)} \, , 
\label{DE2} 
\eea
the NNLO cross section reads:
\bea
\sigma_{NNLO} & = &
\sigma_{S}^{(2,0,\gamma)} \Biggl\{ 1 + \Delta^{(0,Ax)}
+ \sigma_{S}^{(2,1,\gamma)} \Bigl[ 1 +
\Delta^{(0,Ax)} \Bigr] + C_{F} \left( \frac{\alpha_{s}}{2 \pi} \right) 
{\Delta}^{(1,Ax)} \nn\\
& & \hspace{3.5cm} + \sigma_{S}^{(2,2,\gamma)} \Bigl[ 1 +
\Delta^{(0,Ax)} \Bigr] + 
C_{F} \left( \frac{\alpha_{s}}{2 \pi} \right)^2 \Delta^{(2,Ax)} \Biggl\} \,.
\label{finb}
\eea
We now  expand the expression in the curly brackets of
Eq. (\ref{finb}) up to and including terms of order $\beta^0$. Using  
the threshold expansions \cite{us1,us2,us3} of the one- and two-loop
vector and axial vector form factors and putting the renormalization
scale $\mu= m_Q$  we obtain
\bea
\sigma_{NNLO} & = & 
\sigma^{(2,0,\gamma)}_S \Bigl\{ 1 + \Delta^{(0,Ax)}
+ C_{F} \left( \frac{\alpha_{s}}{2 \pi} \right) \Delta^{(1,Ve)} \bigl( 1
+ \Delta^{(0,Ax)} \bigr) + C_{F} \left( \frac{\alpha_{s}}{2 \pi} \right) 
{\Delta}^{(1,Ax)}
\nn\\
& &  
+ C_{F} \left( \frac{\alpha_{s}}{2 \pi} \right)^2  \Delta^{(2,Ve)}
\bigl( 1 + \Delta^{(0,Ax)} \bigr)
+ C_{F} \left( \frac{\alpha_{s}}{2 \pi} \right)^2 
\Delta^{(2 ,Ax)} \Bigr\} \, .
\label{CS}
\eea
The Born cross section   $\sigma^{(2,0,\gamma)}_S$ is given in Eq.~(\ref{ss20g}).
We get for the terms $\Delta^{(1,Ve)}$  and $\Delta^{(2,Ve)}$
which arise from 
 the one- and two-loop photon-exchange contributions, respectively:
\bea
\Delta^{(1,Ve)} & = & \frac{6 \zeta(2)}{\beta} - 8  
+ {\mathcal O} \left( \beta \right) \, , 
\label{1L} \\
\Delta^{(2,Ve)} & = & C_F \Delta^{(2,Ve)}_{A} + C_A \Delta^{(2,Ve)}_{NA} 
+ N_f T_R \Delta^{(2,Ve)}_{L} + T_R \Delta^{(2,Ve)}_{H} \, ,
\eea
with
\bea
\Delta^{(2,Ve)}_{A} & = & \frac{12 \zeta^{2}(2)}{\beta^2} 
- \frac{48 \zeta(2)}{\beta} + 24 \zeta^{2}(2) 
- 4 \zeta(2) \Bigl(  \frac{35}{3} - 8 \ln{2} 
+ 4 \ln{\beta} \Bigr)
- 4 \zeta(3) + 39 
\nn\\
& & 
+ {\mathcal O} \left( \beta \right) \, , 
\label{SS} \\
\Delta^{(2,Ve)}_{NA}  & = & \frac{4 \zeta(2)}{\beta} \Bigl( 
\frac{31}{12} - \frac{11}{2} \ln{(2 \beta)} \Bigr) 
+ 4 \zeta(2) \Bigl( \frac{179}{12} - 16 \ln{2} 
- 6 \ln{\beta} \Bigr)
- 26 \zeta(3) - \frac{151}{9} 
\nn\\
& & 
+ {\mathcal O} \left( \beta \right)\, , \\
\Delta^{(2,Ve)}_{L} & = & \frac{4 \zeta(2)}{\beta} \Bigl( 2 \ln{(2 \beta)}
- \frac{5}{3} \Bigr) + \frac{44}{9}  
+ {\mathcal O} \left( \beta \right)\, , \\
\Delta^{(2,Ve)}_{H} & = & - \frac{32}{3} \zeta(2) + \frac{176}{9} 
+ {\mathcal O} \left( \beta^2 \right)
\, .
\eea
In the equations above, $\zeta(2)=\pi^2/6$,  $C_A=N_c$,  $T_R=1/2$, and
 $N_f$ is the number of light quarks, which we take to be massless. For the
threshold expansions of the terms $\Delta^{(i,Ax)}$ $(i=0,1,2)$, involving
$Z$ boson exchange, we find:
\bea
\Delta^{(0,Ax)}  & = & \frac{s^2}{D_Z} \Biggl\{ 
\frac{\left(\aez\right)^2 + \left(\vez\right)^2}{
\left(\veg \vqg \right)^2} \Biggl[ 2 \left(\aqz\right)^2
\frac{\beta^2}{3-\beta^2} + \left(\vqz\right)^2 \Biggr]
+ 2 \left(1-\frac{m_Z^2}{s}\right) \frac{\vez \vqz}{\veg \vqg} \Biggr\} \, ,
\label{Z0} \\
\Delta^{(1,Ax)} & = &  {\mathcal O} \left( \beta^2 \right) ,
\label{Z1} \\
\Delta^{(2,Ax)} & = &  
\frac{64 \zeta(2) m_Q^4 (a_Q^Z)^2 \left[ (v_e^Z)^2 + (a_e^Z)^2 \right]}{
( v_Q^{\gamma} v_e^{\gamma} )^2 (4m_Q^2 - m_Z^2)^2} C_F 
+ {\mathcal O}(\beta) \, ,
\label{Z2} 
\eea
where $\hv_f$, $\ha_f$ are  defined in Eq. (\ref{vakoppl}).
While the multiplicative factor  $\Delta^{(0,Ax)}$ 
is given in exact form, the higher order terms are expanded to the
appropriate order in $\beta$ such that the NNLO cross section
(\ref{CS})  is obtained to 
next-to-next-to-leading
logarithmic order (NNLL) in $\beta$.
\noindent
We observe:\\
(i) To NNLL in  $\beta$ the NNLO cross section is infrared-finite, i.e., the $Q \bar Q$ cross section
is equal to the  total production cross section to this order. This is due to the fact that
close to threshold, real gluon emission is suppressed by a velocity factor with respect to the
$Q \bar Q$ final state. The two-loop triangle diagrams, studied in \cite{us3}, 
do not contribute to the cross section  to this order in $\beta$.\\
(ii) In the expansions of  $\Delta^{(1,Ve)}$ and $\Delta^{(2,Ve)}$
 IR divergences appear at ${\mathcal O}(\beta)$ and  ${\mathcal
   O}(\beta^2)$, respectively.
In the expansions of $\Delta^{(1,Ax)}$ and $\Delta^{(2,Ax)}$ such
divergences show up to order $\beta^4$ and $\beta^3$, respectively. \\
(iii) The threshold cross section  $\sigma_{NNLO}$ was calculated to
NNLL in
$\beta$ in the above energy region, for pure vector exchange, previously in 
\cite{Czarnecki:1997vz} (see also \cite{Hoang:1997sj}). 
Putting the axial vector couplings to zero in the above expressions and 
comparing with \cite{Czarnecki:1997vz} we find agreement.\\
(iv) The first and second-order $Z$-boson exchange terms
$\Delta^{(1,Ax)}$ and  $\Delta^{(2,Ax)}$, involving the axial
vector coupling $\ha_Q$, are of order $\beta^2$ and $\beta^0$,
respectively. $\Delta^{(2,Ax)}$ is thus  relevant for an analysis
aiming at NNLO  accuracy at threshold (counting $\alpha_s \sim
\beta$).  The analytic result for the axial-vector contribution at 
${\mathcal O}(\alpha_s^2)$, which can be obtained to all orders in 
$\beta$ from  \cite{us2,us3}, has not been  given before in the literature.
The term $\Delta^{(2,Ax)}$ is small compared to $\Delta^{(2,Ve)}$,
see Fig.~\ref{thr}. 

Nevertheless, at a 
high luminosity linear collider with polarized $e^-$ and $e^+$ beams one may 
eventually be able to disentangle the vector and axial-vector induced 
contributions to the $t \bar t$ cross section. A numerical calculation
of axial vector contributions in the context of Lippmann-Schwinger 
equations was performed in  \cite{AX}, and the axial-vector
contribution  to the
threshold cross section at next-to-next-to-leading 
logarithmic order was determined in \cite{Hoang:2001mm} by a
combination
of numerical and analytical calculations.
\par
For completeness we mention that a summation of all terms
of the form $\as^n/\beta^n$ should be performed
when the region of small heavy-quark
velocities $\beta \sim \as$ is approached. The result of the
resummation of these leading terms (and of part
of the subleading terms) is the well-known Sommerfeld-Sakharov factor (see, for
instance \cite{Czarnecki:1997vz}). Subleading terms  may be
resummed by  the renormalization group \cite{Hoang:2001mm} in the
context of effective field theory methods.
In this paper we are not concerned with these summation methods, as we
consider only the region $\beta \gg \as$.

\subsection{Antisymmetric cross cection and  $A_{\mbox{\tiny{FB}}}$}

Next we perform the threshold expansion of the second-order
antisymmetric cross section $\sigma^{(A)}_{NNLO}= \sigma^{(2p)}_{A}$ ,
given in Eqs. (\ref{sas22p}) and  (\ref{sa20ff})   - (\ref{sa22ff}), in 
the same manner as was done above for  $\sigma_{NNLO}$, using the results 
of \cite{us1,us2,us3}. We obtain to NNLL in $\beta$:
\be
\sigma^{(A)}_{NNLO} = \sigma^{(2,0)}_A \Bigl\{ 1
+ C_{F} \left( \frac{\alpha_{s}}{2 \pi} \right) \Delta^{(A,1)} 
+ C_{F} \left( \frac{\alpha_{s}}{2 \pi} \right)^2 \, \Delta^{(A,2)} 
\Bigr\} \, ,
\label{CSA}
\ee
where $\sigma^{(2,0)}_A$ is given in Eq.~(\ref{sa20ff}) and 
\bea
\Delta^{(A,1)} & = & \frac{6 \zeta(2)}{\beta} - 6  
+ {\mathcal O} \left( \beta \right) \, , \nn \\
\Delta^{(A,2)} & = & C_F \Delta^{(A,2)}_{A} + C_A \Delta^{(A,2)}_{NA} 
+ N_f T_R \Delta^{(A,2)}_{L} + T_R  \big( \Delta^{(A,2)}_{H} 
+  \Delta^{(A,2)}_{tr} \big)  \, ,
\label{deldec}
\eea
with
\bea
\Delta^{(A,2)}_{A} & = & 
\frac{12 \zeta^{2}(2)}{\beta^2} 
- \frac{36 \zeta(2)}{\beta} 
+ 24 \zeta^{2}(2) 
- 4 \zeta(2) \Bigl(  \frac{25}{6} 
           - \frac{25}{4} \ln{2} 
           + \frac{9}{2} \ln{\beta} \Bigr)
- \frac{35}{4} \zeta(3) 
+ \frac{70}{3} 
\nn\\
& & 
+ {\mathcal O} \left( \beta \right) \, , \\
\Delta^{(A,2)}_{NA}  & = & \frac{4 \zeta(2)}{\beta} \Bigl( 
\frac{16}{3} 
- \frac{11}{2} \ln{(2 \beta)} \Bigr) 
+ 4 \zeta(2) \Bigl( \frac{67}{6} 
        - \frac{25}{2} \ln{2} 
        - 4 \ln{\beta} \Bigr)
- \frac{35}{2} \zeta(3) 
- 14 
\nn\\
& & 
 + {\mathcal O} \left( \beta \right)\, , \\
\Delta^{(A,2)}_{L} & = & \frac{4 \zeta(2)}{\beta} \Bigl( 2 \ln{(2
  \beta)}        
- \frac{8}{3} \Bigr) + 4
 + {\mathcal O} \left( \beta \right)\, , \\
\Delta^{(A,2)}_{H} & = & - \frac{32}{3} \zeta(2) + \frac{56}{3} 
+ {\mathcal O} \left( \beta^2 \right) \, .
\label{aaapm}
\eea
The term $\Delta^{(A,2)}_{tr}$ in Eq.~(\ref{deldec}) is the contribution of the 
triangle diagrams computed in \cite{us3}. It is infrared- and
ultraviolet-finite.
In the case of $t \bar t$
production it is given by
\be
\Delta^{(A,2)}_{tr}  =  \zeta(2) \left( 16 \ln{2} - \frac{23}{3} \right)
- 8 \ln{2} + \frac{8}{3} \ln^2{2}  + {\mathcal O} \left( \beta^2 \right) \, .
\label{anpm}
\ee
Notice that the the second-order antisymmetric $Q \bar Q$ cross section
(\ref{CSA}) is infrared-finite to NNLL in $\beta$ and is equal to the
total antisymmetric cross section in this order.
The terms  $\Delta^{(A,1)}$ and  $\Delta^{(A,2)}$ become
infrared-divergent to order $\beta^2$ and $\beta$, respectively. 

Finally, the second-order forward-backward asymmetry is given
near threshold  by
\be
A_{\mbox{\tiny{FB}}}^{(Q \bar Q)} =
A_{FB,0} \, C_{FB},
\label{AFBthr}
\ee
where $C_{FB}$ is the ratio of the curly bracket in
Eqs.~(\ref{CSA}) and of the curly bracket in 
(\ref{CS})  divided by $(1+\Delta^{(0,Ax)})$.
To NNLL in $\beta$ it is equal to
the complete  forward-backward asymmetry $A_{\mbox{\tiny{FB}}}^{Q}$.
\par
\begin{figure}[ht]
\begin{center}
\epsfig{file=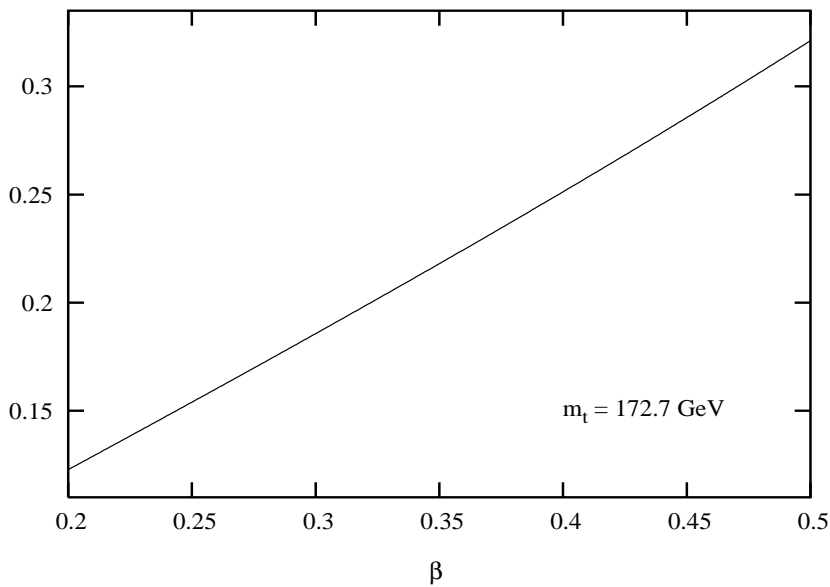, width=12cm, height=8cm}
\end{center}
\caption{\label{AFBthres} The forward-backward
  asymmetry $A_{\mbox{\tiny{FB}}}^{(t \bar t)}$ to NNLL above the
$t \bar t$ threshold in the range  $0.2\leq \beta\leq 0.5$ for $\mu =m_t$.}
\end{figure}

In Fig.~\ref{AFBthres} we have plotted the forward-backward asymmetry
Eq.~(\ref{AFBthr}) to order $\as^2$ for $t\bar t$ production above threshold
in the range $0.2\leq \beta\leq 0.5$, where
$C_{FB}$ is the ratio of two expressions expanded to NNLL in $\beta$. The top mass and the other
parameters were chosen as given in Eq.~(\ref{input}) below. 
A comparison with the exact second order 
asymmetry $A_{\text{\small FB}}^{(t\bar t)}$ will  be made in Section 4.2.

%

%
%
%
%
\section{Numerical Results}

In this section we compute the $Q\bar Q$ contributions to $\afb$ to
order $\as^2$ using the results of Section 2
and the analytic results for the one- and two-loop vector and 
type A and B axial vector form factors given in 
\cite{us1,us2,us3}.
The numerical evaluation of the harmonic polylogarithms that
appear in these expressions were made using the code of \cite{Polylog3}.
For $b$ and $c$ quarks the above formulae 
are evaluated at and in the vicinity of the
$Z$ resonance and for $t$ quarks between $2m_t<\sqrt{s}\leq 1$ TeV.
The quark masses, whose values we use are given below, are defined
in the on-shell scheme while  $\as$ is the QCD coupling in the
$\overline{\rm{MS}}$ scheme. When calculating  $\afb^{(Q\bar Q)}$
for $b$ and $c$ quarks around the $Z$ resonance, $\as$ is defined with
respect to the effective $N_f=5$ flavor theory; that is, the top quark
contribution to the gluon self-energy that enters the vector and
type A axial vector form factors to order $\as^2$ is absent. The
forward-backward asymmetry 
for top quarks is computed in the six-flavor theory with the
corresponding six-flavor QCD 
coupling determined from the five-flavor coupling at the matching point $\mu=m_t$. 
We use the following  input values \cite{:2004qh}:
\bea
m_c= 1.5 \, {\rm GeV}, \quad  m_b= 5 \, {\rm GeV},\quad  m_t= 172.7 \pm
2.9 \, {\rm GeV}, \nn \\
m_Z =91.1875 \, {\rm GeV}, \quad  \Gamma_Z=2.4952\, {\rm GeV}, \nn \\
\sin^2\theta_W=0.23153, \quad \as^{N_f=5}(m_Z)= 0.1187 \,.
\label{input}
\eea
The value of mass of the top quark is the recent CDF and D0 
average \cite{unknown:2005cc}.
For $b$ and $c$ quarks the type B axial vector contributions are
evaluated with the central value of $m_t$ given in~\eqref{input}.
\par
In the following we denote $A_{\mbox{\tiny{FB}}}^{(Q\bar Q)}$
evaluated to order
$\as$ and $\as^2$, respectively, by:
\begin{align}
A_{\mbox{\tiny{FB}}}^{(Q\bar Q)}(\as) = & \; A_{\mbox{\tiny{FB}},0}^{(Q\bar Q)} \left(1 + A_{1}^{(Q\bar Q)}\right),\nn\\
A_{\mbox{\tiny{FB}}}^{(Q\bar Q)}(\as^2) = & \; A_{\mbox{\tiny{FB}},0}^{(Q\bar Q)} \left(1 + A_{1}^{(Q\bar Q)} + A_{2}^{(Q\bar Q,A)} + A_{2}^{(Q\bar Q,B)}\right).\nn
\end{align}
\subsection{$A_{\text{\small FB}}^{(Q\bar Q)}$ for  $b$ and $c$ quarks at and in the vicinity of $\sqrt{s}=m_Z$}

Let us first consider the $b$ quark asymmetry. As it is to be computed for
$\sqrt{s} \simeq m_Z$ we can safely neglect the masses of the $u, $d, $s,$ and $c$ quarks
which contribute to the second order
form factors. (As already
mentioned above, the $t$ quark  contribution to the gluon self-energy
is decoupled.) \\
The type B axial vector form factor $G_1^{(2\ell,B)}$
is non-zero due to the large mass 
splitting between $t$  and $b$ quarks, and to very
good approximation one may neglect in these triangle diagram contributions
the mass of the $b$ quark. Therefore we use
\be
G_1^{(2\ell,B)}(s)=G_1^{(2\ell,B)}(s,m_b=0,m_b=0) - 
G_1^{(2\ell,B)}(s,m_t,m_b=0) \,,
\label{triangelb}
\ee
when evaluating Eq.~\eqref{afb22zb}, respectivley~\eqref{afb22zb1}, for
the $b$ quark. The functions on the right-hand side of~\eqref{triangelb},
whose second and third argument denotes the mass of the quark in the loop 
and the mass of the external quark, respectively, are given in
\cite{Kniehl:1989qu,us3}. 
(We use the notation of \cite{us3}.) 
Putting $\beta=1$ in Eq.~\eqref{afb22zb1}, our result for 
$A_{\mbox{\tiny{FB}}}^{(2p,B)}(s=m_Z^2)$ agrees with that of \cite{Altarelli}. \\
Table~\ref{tb:afbz} contains the values for the lowest order forward-backward asymmetry at the
$Z$ resonance, together with the $b \bar b$ contributions to first and second order
in $\as$ for three choices of the renormalization
scale $\mu$.  The  photon and $\gamma-Z$ interference contributions, which are (on the
$Z$ resonance) of higher order in the electroweak
couplings, are not taken into account. Table~\ref{tb:afbz} shows that the QCD corrections are dominated
by the type B contributions: they are about three times as large as the order $\as$ and about nine times
as large as  the order $\as^2$ corrections. This is due to the fact that
we are close to the
chiral limit,
as $m_b/m_Z \ll 1$.
In this limit the order $\as$ vector and axial vector
form factors $F_1^{(1\ell)},$ $G_1^{(1\ell)},$
and the  order $\as^2$
vector and type A axial vector form factors $F_1^{(2\ell)},$
$G_1^{(2\ell, A)}$ become equal while the chirality-flipping 
form factors vanish. In this limit  $A_{1}^{(Q\bar Q)}$ and
$A_{2}^{(Q{\bar Q}, A)}$ vanish, too,  as an inspection of the above formulae
shows.
Thus the QCD corrections
are dominated by the type B contribution.
As it turns out, it amounts to
a correction of the lowest order asymmetry by only about one per mille. 
\par
\begin{figure}[ht]
\begin{center}
\epsfig{file=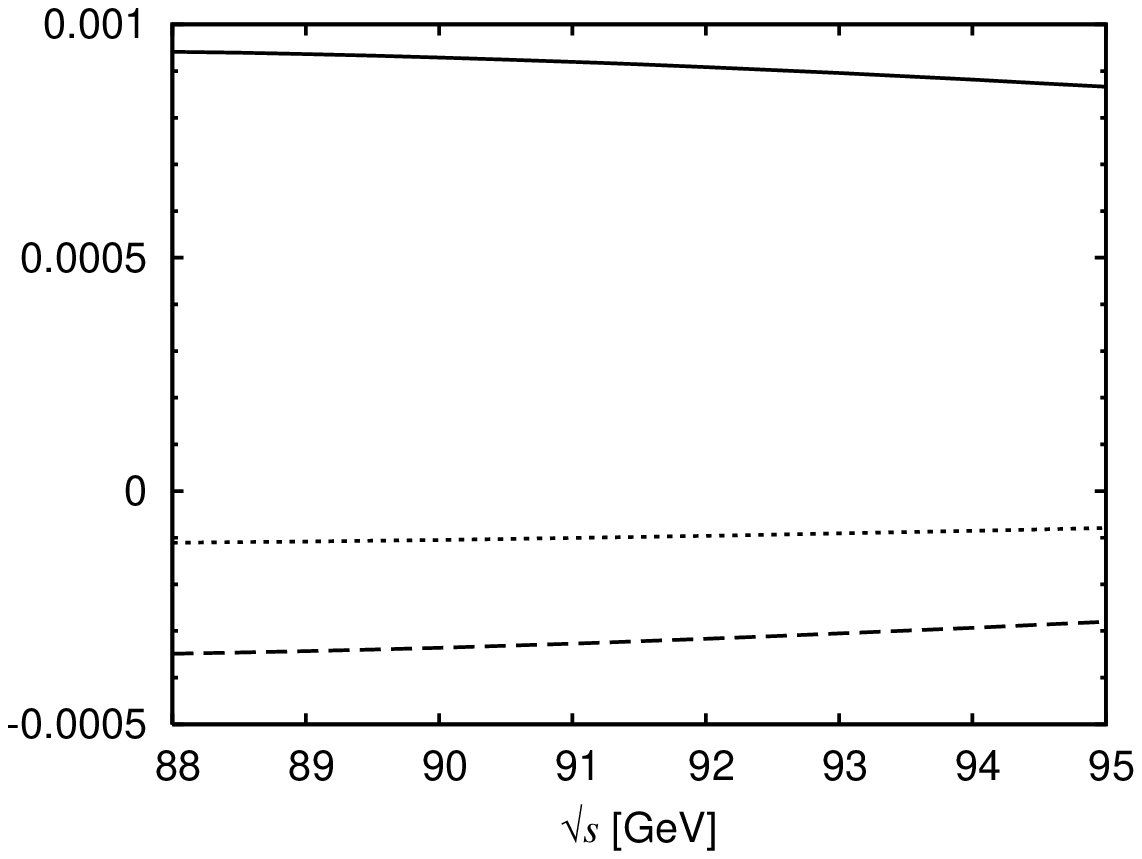, width=12cm, height=8cm}
\end{center}
\caption{First and second order QCD corrections  $A_{1}^{(b\bar b)}$
 (dashed),  $A_{2}^{(b\bar b,A)}$ (dotted), and  $A_{2}^{(b\bar b,B)}$
(solid) for $\mu = m_Z$ in the vicinity of the $Z$ resonance.}\label{fig:2b}
\end{figure}

In Fig.~\ref{fig:2b}  the  first and second order QCD corrections  $A_{1}^{(b\bar b)}$,
 $A_{2}^{(b\bar b,A)}$, and  $A_{2}^{(b\bar b,B)}$, evaluated for 
 $\mu = m_Z$, are shown between 88 GeV $<\sqrt{s}<$ 95 GeV. Here the
contributions from photon exchange are included.  Again the QCD corrections are dominated by the
type B triangle diagram contributions. Varying the
renormalization scale in the range  $m_Z/2 \leq \mu \leq 2 m_Z$ changes these
numbers only by a small amount.
 
\begin{table}
\centering
\begin{tabular}{|c||c|c|c|c|c|c|}\hline
& $A_{\mbox{\tiny{FB}},0}^{(b\bar{b})}$ & $A_{1}^{(b\bar{b})}$ & $A_{2}^{(b\bar{b},A)}$ & $A_{2}^{(b\bar{b},B)}$ & $A_{\mbox{\tiny{FB}}}^{(b\bar{b})}(\as)$ &  $A_{\mbox{\tiny{FB}}}^{(b\bar{b})}(\as^2)$ \\ \hline\hline
$\mu=\frac{m_Z}{2}$ & 0.103128 & -0.000365 & -0.000084 &  0.001147 & 0.103090 & 0.103200 \\\hline
$\mu=m_Z$           & 0.103128 & -0.000326 & -0.000100 &  0.000919 & 0.103094 & 0.103179 \\\hline
$\mu= 2 m_Z $       & 0.103128 & -0.000295 & -0.000109 &  0.000753 & 0.103097 & 0.103164 \\\hline
\end{tabular}
\caption{\label{tb:afbz} The $b \bar b$ contributions to $\afb$ for bottom quarks at $\sqrt{s}=m_Z$.}
\end{table}

Next we  consider the $c$ quark asymmetry for $\sqrt{s}\simeq m_Z$. To very good approximation we can neglect
the masses of the $c$ and $b$ quarks in their contribution to the gluon self-energy. 
Here the type B axial vector form factor $G_1^{(2\ell,B)}$ is again
determined by  the large mass splitting between $t$  and $b$ quarks. In view of the convention adopted 
in Eq.~\eqref{decomp}, where the neutral current couplings of the
external quark are factored out, 
we use now 
\be
G_1^{(2\ell,B)}(s)= G_1^{(2\ell,B)}(s,m_t,m_c=0) - G_1^{(2\ell,B)}(s,m_b=0,m_c=0) 
\label{triangelc}
\ee
when applying Eq.~\eqref{afb22zb}, respectively~\eqref{afb22zb1}, to the 
$c$ quark. Again we put $\beta=1$ in these
equations. Eq.~\eqref{triangelc}
is equal in magnitude but opposite in sign to Eq.~\eqref{triangelb}. \\
Table~\ref{tc:afbz} contains the values for the lowest order $c$ quark forward-backward asymmetry at the
$Z$ resonance -- without the $\gamma$ and $\gamma-Z$ contributions  --,
together with the $c \bar c$ contributions 
to first and second order
in $\as$ for three choices of the renormalization
scale $\mu$. The QCD corrections are dominated
again by the type B term, which in this case is  about two per mille
of the leading order asymmetry. The increase by a factor of about two
as compared to the $b$ quark results from the fact that
$|v_c^Z|<|v_b^Z|$, c.f. Eq.~\eqref{afb22zb1}.

\begin{figure}[]
\begin{center}
\epsfig{file=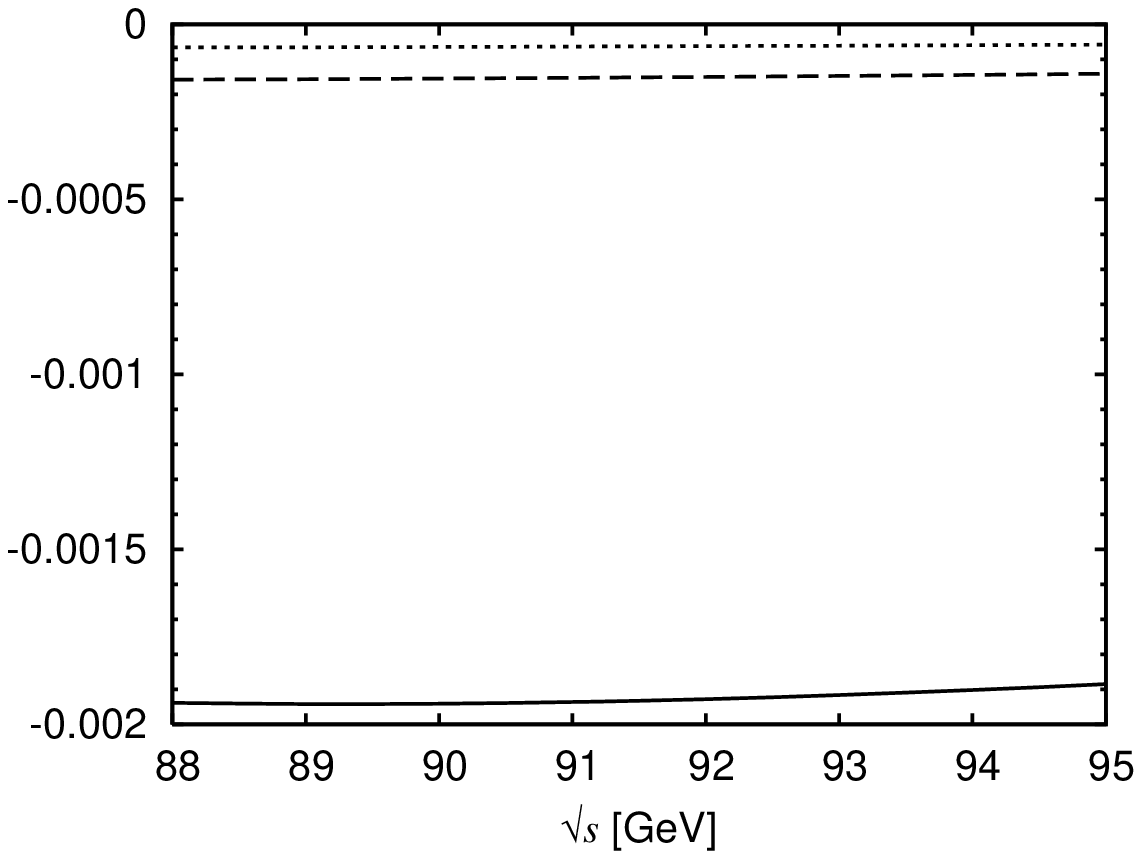, width=12cm, height=8cm}
\end{center}
\caption{First and second order QCD corrections  $A_{1}^{(c\bar c)}$
 (dashed),  $A_{2}^{(c\bar c,A)}$ (dotted), and  $A_{2}^{(c\bar c,B)}$
(solid)  for $\mu = m_Z$ in the vicinity of the $Z$ resonance.}\label{fig:2c}
\end{figure}

In Fig.~\ref{fig:2c}  the  first and second order QCD corrections  $A_{1}^{(c\bar c)}$,
 $A_{2}^{(c\bar c,A)}$, and  $A_{2}^{(c\bar c,B)}$, evaluated for 
 $\mu = m_Z$, are shown between 88 GeV $<\sqrt{s}<$ 95 GeV,
including the contributions from photon exchange.  For $c$ quarks,
too, 
the QCD corrections are dominated by the
type B triangle diagram contributions. 

\begin{table}
\centering
\begin{tabular}{|c||c|c|c|c|c|c|}\hline
& $A_{\mbox{\tiny{FB}},0}^{(c\bar{c})}$ & $A_{1}^{(c\bar{c})}$ & $A_{2}^{(c\bar{c},A)}$ & $A_{2}^{(c\bar{c},B)}$ & $A_{\mbox{\tiny{FB}}}^{(c\bar{c})}(\as)$ &  $A_{\mbox{\tiny{FB}}}^{(c\bar{c})}(\as^2)$ \\ \hline\hline
$\mu=\frac{m_Z}{2}$ & 0.073592 & -0.000170 & -0.000060 &  -0.002418 & 0.073580 & 0.073397 \\\hline
$\mu=m_Z$           & 0.073592 & -0.000152 & -0.000063 &  -0.001938 & 0.073581 & 0.073434 \\\hline
$\mu= 2 m_Z $       & 0.073592 & -0.000138 & -0.000064 &  -0.001586 & 0.073582 & 0.073460 \\\hline
\end{tabular}
\caption{\label{tc:afbz} The $c \bar c$ contributions to $\afb$ for charm quarks at $\sqrt{s}=m_Z$.}
\end{table}

\begin{figure}[]
\begin{center}
\epsfig{file=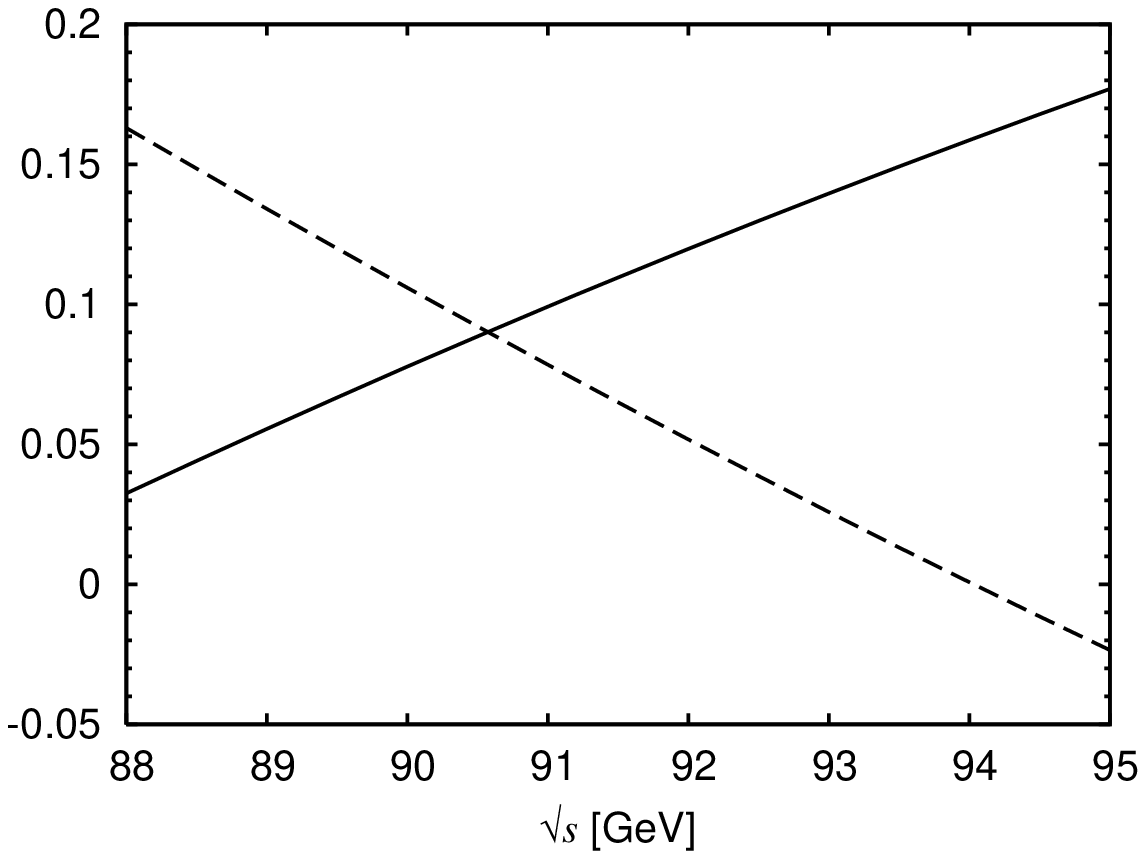, width=12cm, height=8cm}
\end{center}
\caption{$A_{\mbox{\tiny{FB}}}^{(b\bar b)}(\as^2)$ (solid)
and $A_{\mbox{\tiny{FB}}}^{(c\bar c)}(\as^2)$ (dashed)
for  $\mu = m_Z$ in the vicinity of the $Z$ resonance.}\label{fig:2bc}
\end{figure}
In Fig.~\ref{fig:2bc} the $b \bar b$ and $c\bar c$  forward-backward
 asymmetries
to order $\as^2$, $A_{\mbox{\tiny{FB}}}^{(b\bar{b})}(\as^2)$ and $A_{\mbox{\tiny{FB}}}^{(c\bar{c})}(\as^2)$ 
are displayed between 88 GeV $<\sqrt{s}<$ 95 GeV, using $\mu =m_Z$.
Figs.~\ref{fig:2b},~\ref{fig:2c} show that the  order $\as^2$ asymmetry is
increased, in the case of $b$ quarks,  by about one per mille and
decreased, in the case of $c$ quarks, by about two per mille of the
respective lowest order asymmetry in the whole energy
range considered. Varying the
renormalization scale in the range  $m_Z/2 \leq \mu \leq 2 m_Z$ changes
the asymmetries shown in Fig.~\ref{fig:2bc} only by a very small
amount. 
\par 
The complete forward-backward asymmetry to order $\as^2$ for $b$ and
$c$ quarks at and in the vicinity of
the $Z$ resonance is dominated by the respective three- and four-parton contributions. A complete computation
of these terms to order $\as^2$ has not yet been done for $m_b, m_c
\neq 0$. Nevertheless,  we expect that the respective results of  
\cite{seymour}, which were obtained in the
massless limit, will not change dramatically.

\subsection{$A_{\text{\small FB}}^{(t\bar t)}$ for top  quarks above threshold}

Finally we compute the $t \bar t$ contributions to the
forward-backward asymmetry in the reaction $e^+ e^-\to t {\bar t}X$,
sufficiently far away from the pair production threshold in order that
perturbation theory in $\as$ is applicable. That is, the following
results apply to events with $t$ and $\bar t$ velocities $\beta \gg
\as$.  As already stated above, $\as$ is defined in the 
 six flavor QCD, with all quarks but the top quark taken to be
massless. The  value of $\as(\mu=m_t)$ is determined from the input
value~\eqref{input}, and values of $\as$ at other energy scales
are obtained by two-loop
renormalization group evolution. Most of the results below are
presented for three values of the top quark mass: the present central
 and 1 s.d. upper and lower values 172.7 GeV, 175.6 GeV, and 169.8 GeV,
respectively.
The type B contribution to $A_{{\small FB}}^{(t\bar t)}$, Eq.~\eqref{afb22zb},
is computed with the two-loop 
axial vector form factor 
\be
G_1^{(2\ell,B)}(s)= G_1^{(2\ell,B)}(s,m_t,m_t) -
G_1^{(2\ell,B)}(s,m_b=0,m_t) \, .
\label{triangelt}
\ee
which is given in  \cite{us3}. 
\begin{figure}[]
\begin{center}
\epsfig{file=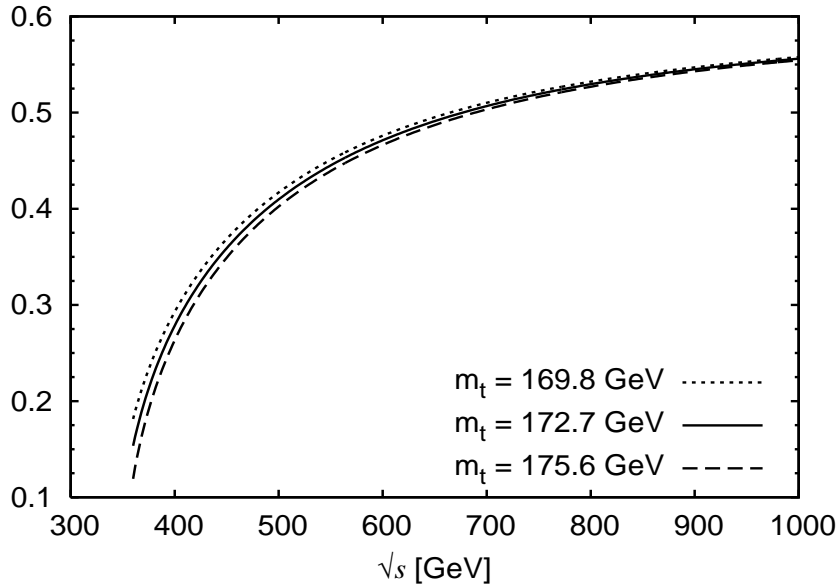, width=12cm, height=8cm}
\end{center}
\caption{Leading order asymmetry $A_{\mbox{\tiny{FB}},0}^{(t\bar t)}$
for three values of the top quark mass.}\label{fig:t0}
\end{figure}

\begin{figure}[]
\begin{center}
\epsfig{file=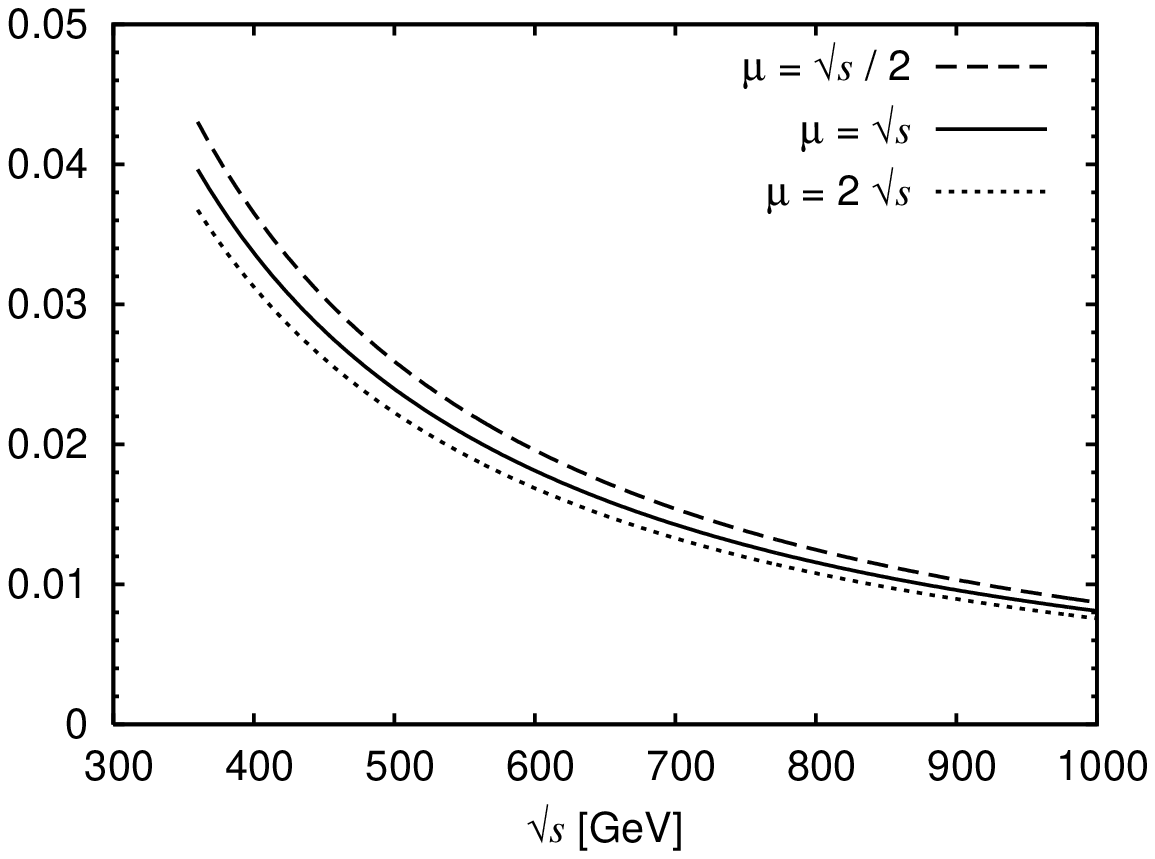, width=12cm, height=8cm}
\end{center}
\caption{Order $\as$ correction  $A_{1}^{(t \bar t)}$
for three values of the renormalization scale $\mu$, using
$m_t=172.7$ GeV.}\label{fig:t1ren}
\end{figure}

\begin{figure}[]
\begin{center}
\epsfig{file=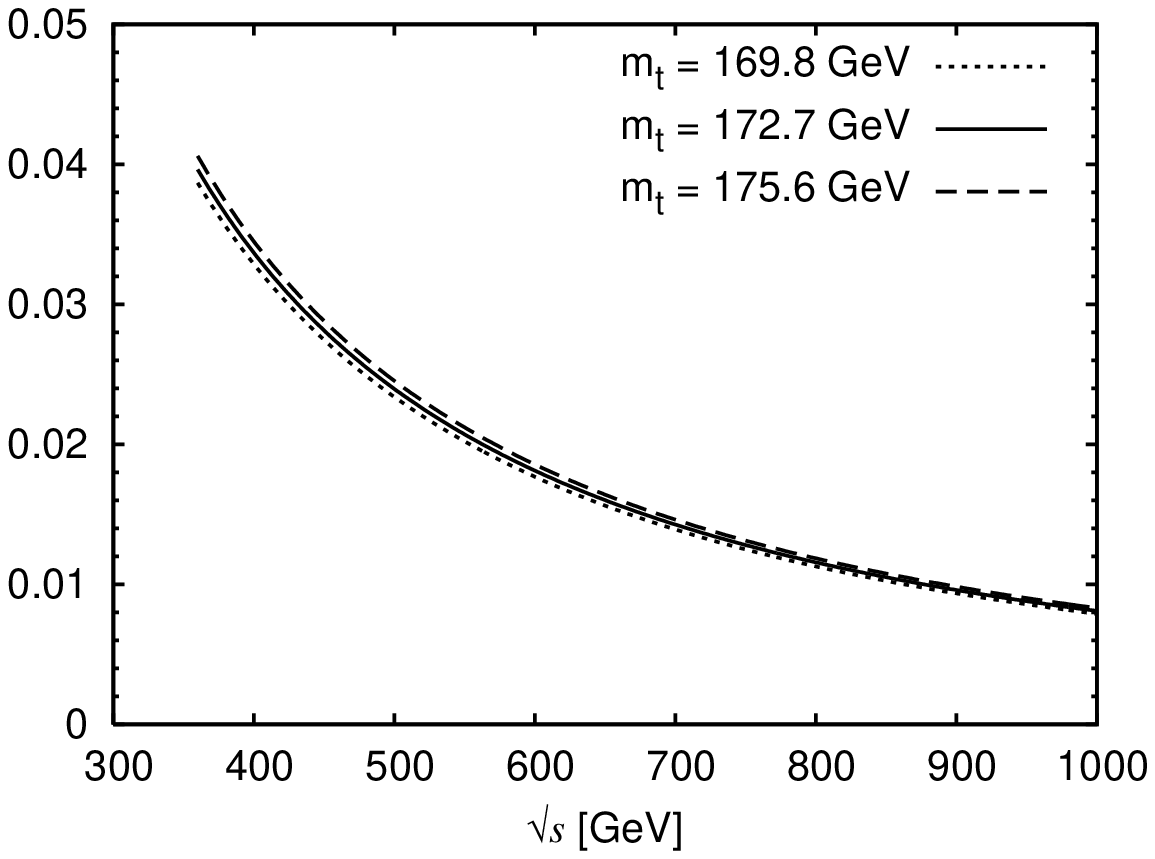, width=12cm, height=8cm}
\end{center}
\caption{Order $\as$ correction  $A_{1}^{(t\bar t)}$
for three values of the top quark mass, using $\mu=\sqrt{s}$.}
\label{fig:t1mass}
\end{figure}

\begin{figure}[]
\begin{center}
\epsfig{file=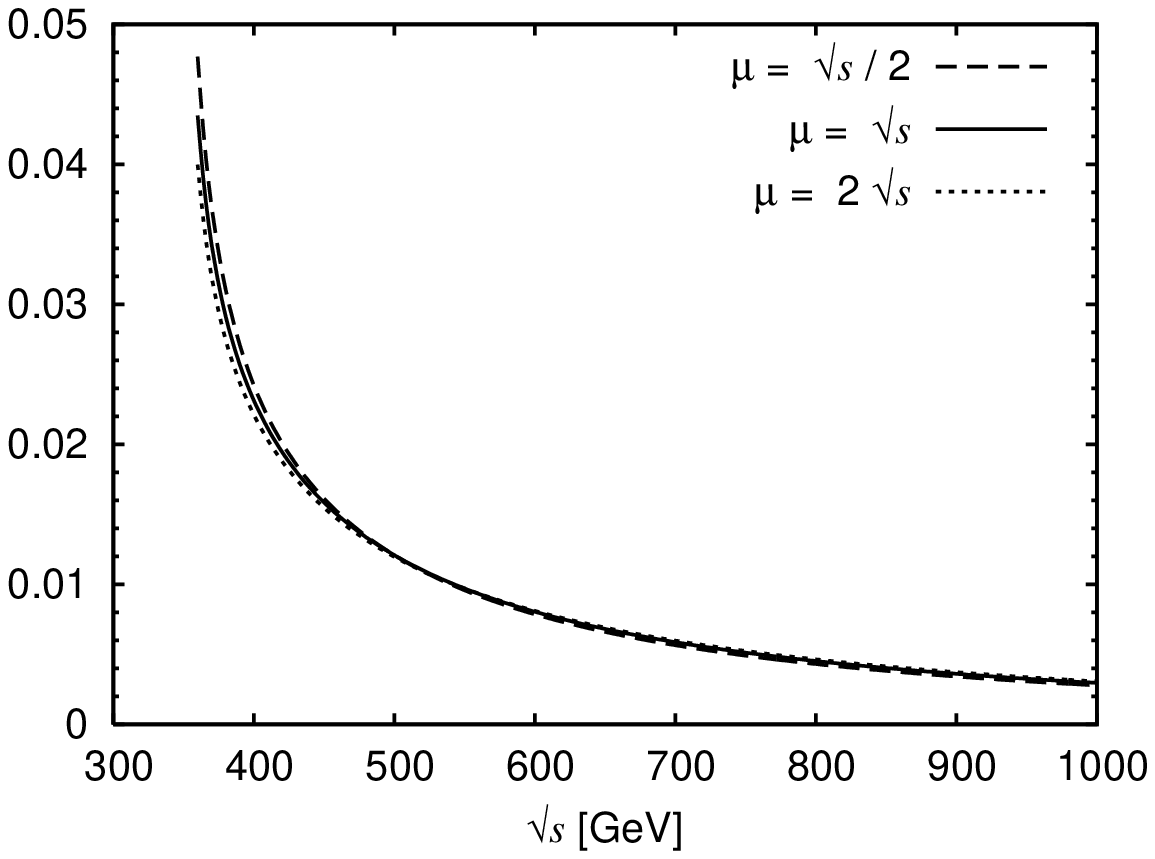, width=12cm, height=8cm}
\end{center}
\caption{Order $\as^2$ correction  $A_{2}^{(t\bar t,A)}$
for three values of the renormalization scale $\mu$, using
$m_t=172.7$ GeV.}\label{fig:t2aren}
\end{figure}

\begin{figure}[]
\begin{center}
\epsfig{file=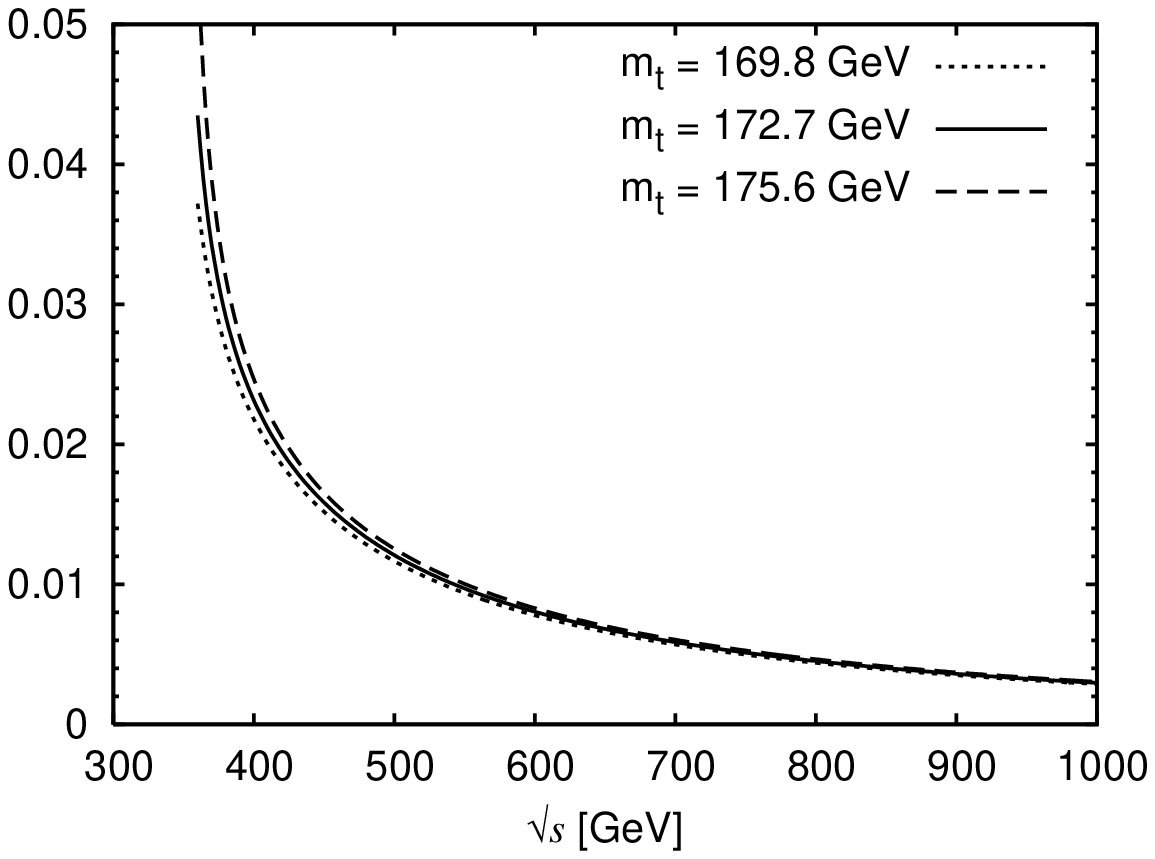, width=12cm, height=8cm}
\end{center}
\caption{Order $\as^2$ correction  $A_{2}^{(t\bar t,A)}$
for three values of the top quark mass, using $\mu=\sqrt{s}$.}
\label{fig:t2amass}
\end{figure}

\begin{figure}[]
\begin{center}
\epsfig{file=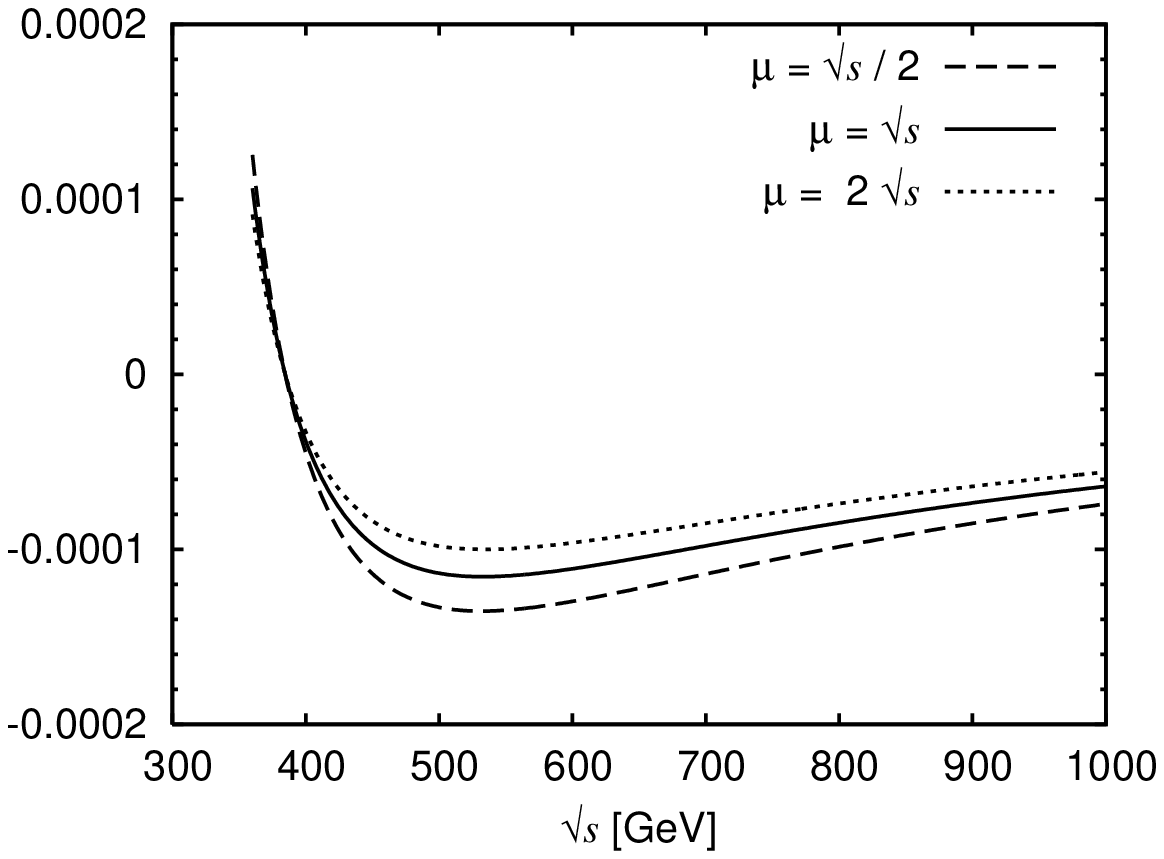, width=12cm, height=8cm}
\end{center}
\caption{Order $\as^2$ correction  $A_{2}^{(t\bar t,B)}$
for three values of the renormalization scale $\mu$, using
$m_t=172.7$ GeV.}\label{fig:t2bren}
\end{figure}

\begin{figure}[]
\begin{center}
\epsfig{file=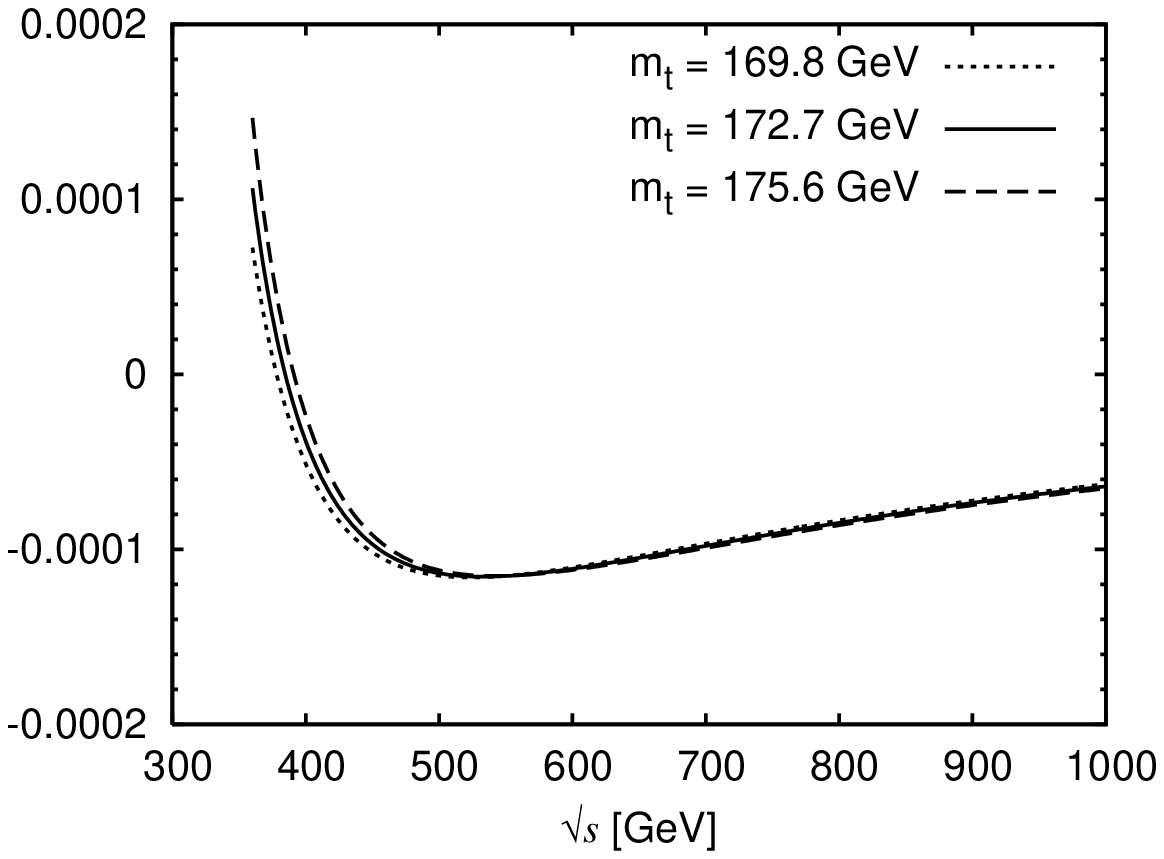, width=12cm, height=8cm}
\end{center}
\caption{Order $\as^2$ correction  $A_{2}^{(t\bar t,B)}$
for three values of the top quark mass, using $\mu=\sqrt{s}$.}
\label{fig:t2bmass}
\end{figure}

\begin{figure}[]
\begin{center}
\epsfig{file=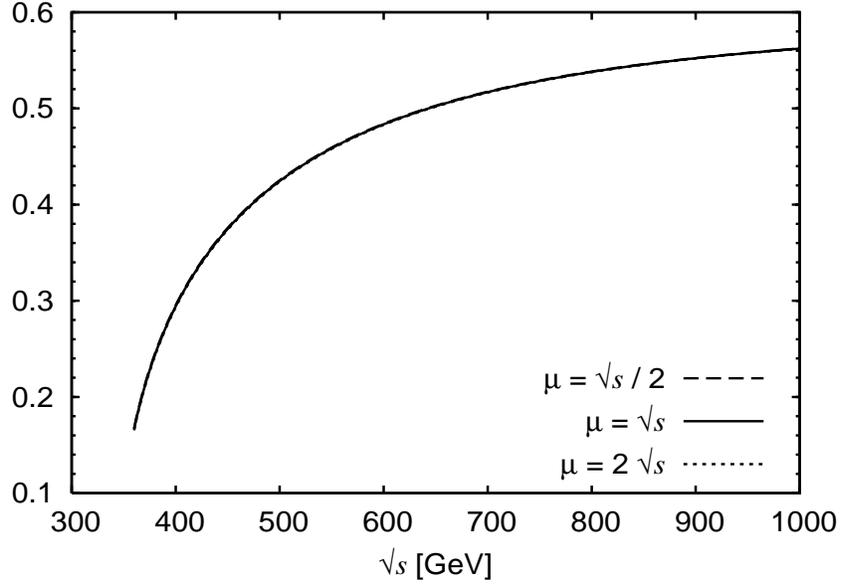, width=12cm, height=8cm}
\end{center}
\caption{Forward-backward asymmetry $A_{\mbox{\tiny{FB}}}^{(t\bar t)}(\as^2)$ 
for three values of the renormalization scale $\mu$, using
$m_t=172.7$ GeV.}\label{fig:tg2ren}
\end{figure}

\begin{figure}[]
\begin{center}
\epsfig{file=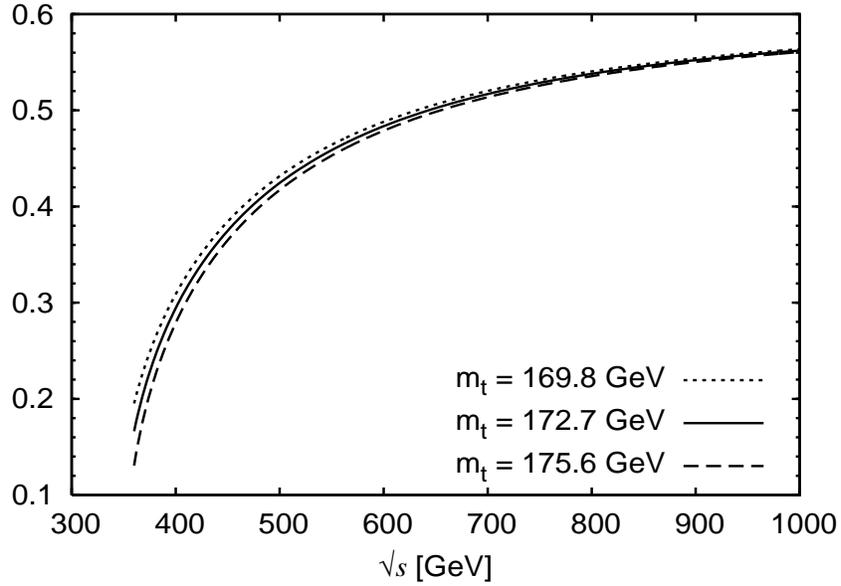, width=12cm, height=8cm}
\end{center}
\caption{Forward-backward asymmetry $A_{\mbox{\tiny{FB}}}^{(t\bar t)}(\as^2)$ 
for three values of the top quark mass, using $\mu=\sqrt{s}$.}
\label{fig:tg2mass}
\end{figure}

\begin{figure}[]
\begin{center}
\epsfig{file=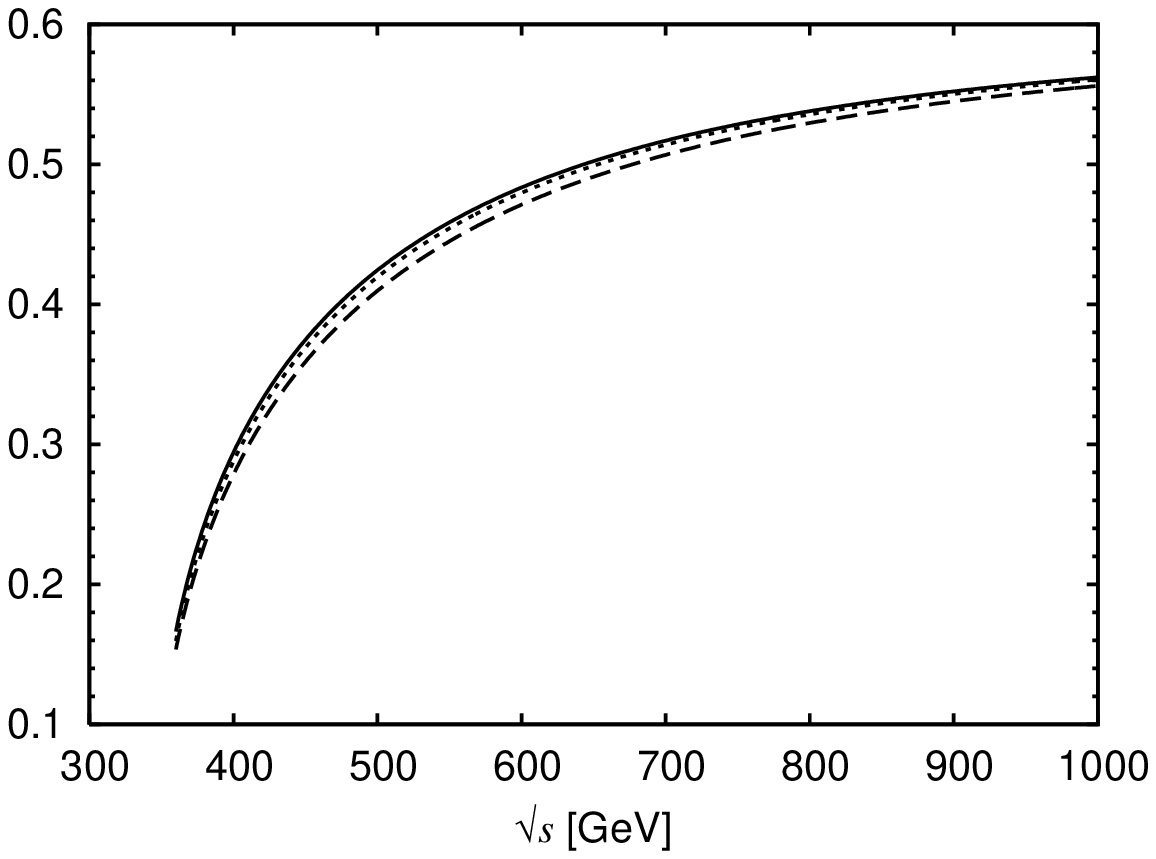, width=12cm, height=8cm}
\end{center}
\caption{Forward-backward asymmetry to lowest, first and second order
in $\as$ using $m_t=172.7$ GeV and
$\mu=\sqrt{s}$. $A_{\mbox{\tiny{FB}},0}^{(t\bar t)}$ (dashed),
$A_{\mbox{\tiny{FB}}}^{(t\bar t)}(\as)$ (dotted), 
$A_{\mbox{\tiny{FB}}}^{(t\bar t)}(\as^2)$ (solid).}
\label{fig:tg012}
\end{figure}
In the following we consider the energy range 360 GeV $\leq \sqrt{s}
\leq$ 1 TeV. In Fig.~\ref{fig:t0}
the leading order asymmetry $A_{\mbox{\tiny{FB}},0}^{(t\bar t)}$
is shown for three values of the top quark mass.
In Figs.~\ref{fig:t1ren} and~\ref{fig:t1mass} the 
order $\as$ correction  $A_{1}^{(t\bar t)}$ is displayed 
for three values of the renormalization scale $\mu$ and fixed top
quark mass, and for three values of $m_t$ and fixed $\mu$,
respectively. The anologous cases are shown in  Figs.~\ref{fig:t2aren}
and~\ref{fig:t2amass} for the
order $\as^2$ correction  $A_{2}^{(t\bar t,A)}$. The 
triangle diagram contributions $A_{2}^{(t\bar t,B)}$
are given in  Fig.~\ref{fig:t2bren}  and
Fig.~\ref{fig:t2bmass} for three values of $\mu$ and $m_t$, respectively.
From these figures we conclude that the two-parton QCD corrections to the
lowest order asymmetry are  moderate to small  for $\sqrt{s}\geq$ 400 GeV.
At $\sqrt{s} =$ 400 GeV, $A_{1}^{(t\bar t)}$ is about 3.3  percent
while $A_{2}^{(t\bar t,A)}$ is about 2.4 percent.
As expected, the relative
importance of the order $\as^2$ corrections increases as the centre-of-mass
energy approaches the threshold region: for $\sqrt{s}\simeq$ 360
GeV, $A_{2}^{(t\bar t,A)}$ is larger than  $A_{1}^{(t\bar t)}$,
signaling that perturbation theory in $\as$ is no longer applicable.
Contrary to the case of $b$ and $c$ quarks at the
$Z$ resonance the two-loop type B contributions are two orders
of magnitude smaller than  $A_{2}^{(t\bar t,A)}$.
\par
Figs.~\ref{fig:tg2ren} and~\ref{fig:tg2mass} show the
forward-backward asymmetry $A_{\mbox{\tiny{FB}}}^{(t\bar t)}(\as^2)$ 
for three values of the renormalization scale and three values of the
top quark mass, respectively. The dependence of the second order
asymmetry on $\mu$ is small: changing $\mu$ from $\sqrt{s}/2$ to
$2\sqrt{s}$ changes $A_{\mbox{\tiny{FB}}}^{(t\bar t)}(\as^2)$, for
fixed $m_t$, only by about 1 percent at $\sqrt{s}\gtrsim$ 360 GeV, and
this dependence on $\mu$ decreases with increasing  c. m. energy.
\par
In Fig.~\ref{fig:tg012} the $t \bar t$ asymmetry is displayed
to lowest, first, and second order in $\as$. 
This figure shows that for c. m. energies sufficiently away from
threshold  the QCD corrections are under control.

\begin{figure}[ht]
\begin{center}
\epsfig{file=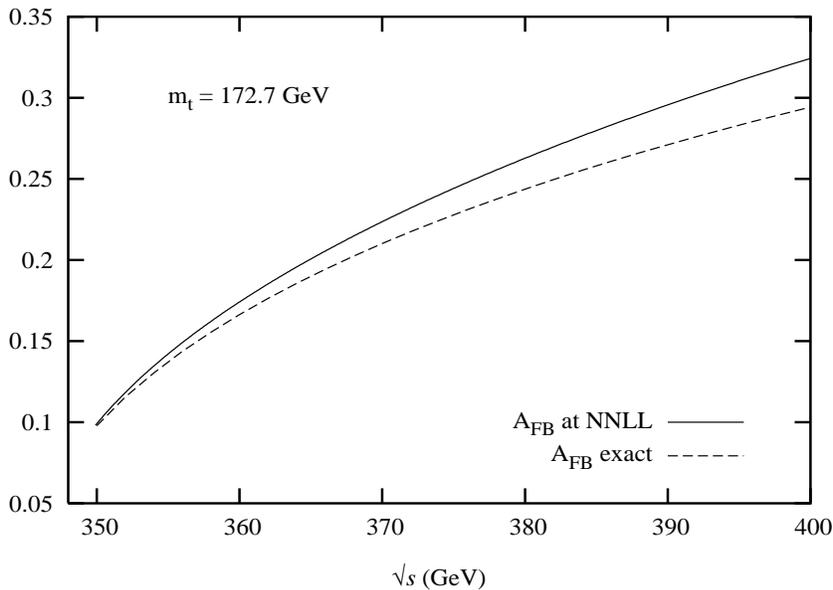, width=12cm, height=8cm}
\end{center}
\caption{The second order forward-backward asymmetry $A_{\mbox{\tiny{FB}}}^{(t\bar t)}(\as^2)$
near threshold: exact values (dashed) as given in
Fig.~\ref{fig:tg012} and the values obtained from the near-threshold
formula Eq.~(\ref{AFBthr}) (solid), using $\mu=m_t=172.7$ GeV.}
\label{fig:trex}
\end{figure}

Finally, a comparison is made in Fig.~\ref{fig:trex} 
between the exact second order forward-backward asymmetry 
$A_{\mbox{\tiny{FB}}}^{(t\bar t)}(\as^2)$  as given in
Fig.~\ref{fig:tg012} and the values obtained from the near-threshold
NNLL formula Eq.~(\ref{AFBthr}). For $\sqrt{s} \lesssim$ 360 GeV
corresponding to $\beta \lesssim 0.3$ the deviation of the NNLL from
the respective exact
value is less than 5 percent.
%
%
%
%
%
\section{Summary and Outlook}

Future experiments on heavy-quark production 
at a planned linear $e^+ e^-$ collider aim at  very precise
measurements of the neutral-current couplings of these quarks. 
An important observable for this purposes  is the forward-backward asymmetry $A_{FB}^Q$
in inclusive heavy quark production, $e^+ e^- \to \gamma^*, Z^* \to Q +X$. The projected
accuracies with which $A_{FB}^Q$ can be measured in future $b$ or $t$
quark production requires also the precise determination of these observables
within the Standard Model. In particular, a computation of the order
$\as^2$ QCD contributions to  $A_{FB}^Q$ for massive quarks is
mandatory.

In view of these perspectives we have calculated 
the contribution of the $Q \bar Q$ final state to  $A_{FB}^Q$ in NNLO
QCD. As discussed above, and explicitly shown for the  $Q \bar Q$
final state, the contributions of the two-parton and of the
three- plus four-parton states to the second-order
forward-backward asymmetry are separately infrared-finite. 
We have provided formulae for the symmetric and antisymmetric $Q \bar
Q$ cross sections  $\sigma_S$ and $\sigma_A$ which yield $A_{FB}^{(Q \bar Q)}$.
These formulae hold for any center-of-mass energy.
Specifically,  in the energy region near threshold where the quark 
velocity $\beta$ satisfies $\alpha_s \ll \beta \ll 1$, we have
expanded  the order $\as$ and $\as^2$  QCD corrections
to $\sigma_S$ and $\sigma_A$  to NNLL in  $\beta$. To this order in
$\beta$ the $Q \bar Q$ cross sections are equal
to the corresponding total cross sections. Therefore,  an (analytic)
expression is obtained for 
the forward-backward asymmetry  $A_{FB}^Q$  to order $\as^2$ and order
$\beta$ near threshold. 

Moreover, we have computed  
the two-parton forward-backward asymmetry $A_{FB}^{(Q \bar Q)}$ 
for $b$ and $c$ quarks on and near the $Z$-boson resonance and for $t$ quarks
for  center-of-mass energies  $\sqrt s$ above threshold to 1 TeV. The two-parton
asymmetry is  determined by the heavy-quark vector and axial vector form factors. 
To order $\as^2$ the axial vector form factors receive besides type A
(universal) corrections also triangle diagram contributions resulting
from the large mass splitting between $t$ and $b$ quarks. These
triangle diagram terms dominate,  for $\sqrt s \sim m_Z$, the QCD corrections from the $Q \bar
Q$ final state  to 
 $A_{FB}^b$ and $A_{FB}^c$. This is due to the
 fact that here  one is close to the chiral limit. However, the
 complete order $\as^2$ QCD corrections
 to these asymmetries are dominated by the contributions from the
 three- and four-parton final states, which were  calculated so far
 only for massless quarks \cite{seymour}.
For top quarks the triangle diagram contributions to $A_{FB}^t$ 
are negligible compared to the type A corrections. These corrections  from the $t \bar t$ final state
to the lowest order aymmetry are
moderate for large  $\sqrt s$ and increase in size towards
threshold. The order $\as^2$ corrections are important, as the analysis
in Section 4.2 shows.

We plan to determine in the near future also the contribution of the
three- and four parton-final states to the  order $\as^2$
forward-backward asymmetry for massive quarks.

%
%
%
%
\section*{Acknowledgment}
This work was  supported 
by Deutsche Forschungsgemeinschaft (DFG),
SFB/TR9, by DFG-Graduiertenkolleg RWTH Aachen, 
by the Swiss
National Science Foundation (SNF) under contract 200020-109162,
by the European Union under the 
contract HPRN-CT2002-00311 (EURIDICE),
by a European Commission Marie Curie Fellowship under
contract number MEIF-CT-2006-024178,
by MCYT (Spain) under Grant FPA2004-00996, by Generalitat Valenciana (Grants 
GRUPOS03/013 and GV05/015), and  by the USA DoE under
the grant DE-FG03-91ER40662, Task J.
%
%
%
%

%
%

\begin{thebibliography}{99}
\def    \np     #1#2#3{{\it Nucl. Phys.} {\bf #1} (19#2) #3}
\def    \nptwoth     #1#2#3{{\it Nucl. Phys.} {\bf #1} (20#2) #3}
\def    \prep   #1#2#3{{\it Phys. Rep.} {\bf #1}  (19#2) #3}
\def    \pl     #1#2#3{{\it Phys. Lett.} {\bf #1} (19#2) #3}
\def    \pltwoth     #1#2#3{{\it Phys. Lett.} {\bf #1} (20#2) #3}
\def    \plold  #1#2#3{{\it Phys. Lett.} {\bf B#1} (19#2) #3}
\def    \prl    #1#2#3{{\it Phys. Rev. Lett.} {\bf #1}  (19#2) #3}
\def    \prltwoth    #1#2#3{{\it Phys. Rev. Lett.} {\bf #1}  (20#2) #3}
\def    \pr     #1#2#3{{\it Phys. Rev.} {\bf #1}  (19#2) #3}
\def    \prd    #1#2#3{{\it Phys. Rev.} {\bf D#1}  (19#2) #3}
\def    \prdtwoth    #1#2#3{{\it Phys. Rev.} {\bf D#1}  (20#2) #3}
\def    \zeit   #1#2#3{{\it Z. Phys.} {\bf C#1}  (19#2) #3}
\def    \cmp    #1#2#3{{\it Comm. Math. Phys.} {\bf #1}  (19#2) #3}
\def    \ibid   #1#2#3{{\it ibid.} {\bf #1} (19#2) #3}
\def    \nc     #1#2#3{{\it Nuovo Cim.} {\bf #1} (19#2) #3}
\def    \acta   #1#2#3{{\it Acta Phys. Polon.} {\bf #1} (19#2) #3}
\def    \tmp    #1#2#3{{\it Theor. Math. Phys.} {\bf #1} (19#2) #3}
\def    \comp    #1#2#3{{\it Comput. Phys. Commun.} {\bf #1} (20#2) #3}
\def    \hepph  #1 {{\tt hep-ph/#1}}
\def    \hepex  #1 {{\tt hep-ex/#1}}
\def    \mathph  #1 {{\tt math-ph/#1}}
\parskip 0pt
\itemsep=0pt



\bibitem{:2004qh}
    [LEP and SLD Collaborations], 
  ``Precision electroweak measurements on the Z resonance,''
  \hepex{0509008}.

\bibitem{Abbaneo:1998xt}
  D.~Abbaneo {\it et al.}  [LEP Heavy Flavor Working Group],
  Eur.\ Phys.\ J.\ C {\bf 4} (1998) 185. \\
LEP/SLD heavy flavour working group, ``Final Input Parameters
for the LEP/SLD Heavy Flavour Analyses'', LEPHF/2001-01;
{\tt http://lepewwg.web.cern.ch/LEPEWWG/heavy/}

\bibitem{tesla}
 J.~A.~Aguilar-Saavedra {\it et al.}  [ECFA/DESY LC Physics Working Group
 Collaboration], ``TESLA Technical Design Report Part III: Physics at an 
$e^+e^-$ Linear Collider'', DESY-report 2001-011 (\hepph{0106315}).



\bibitem{Hawkings:1999ac}
  R.~Hawkings and K.~M\"onig,
  Eur.\ Phys.\ J.\ direct C {\bf 1} (1999) 8
  [\hepex{9910022}].

\bibitem{Erler:2000jg}
  J.~Erler, S.~Heinemeyer, W.~Hollik, G.~Weiglein and P.~M.~Zerwas,
  Phys.\ Lett.\ B {\bf 486} (2000) 125
  [\hepph{0005024}].


\bibitem{Bohm:1989pb}
  M.~B\"ohm {\it et al.},
  ``Forward - Backward Asymmetries,''
in: CERN Yellow Report ``$Z$ Physics at LEP 1'', 
 CERN 89-08 (1989), G. Altarelli et al. (eds.).


\bibitem{Bardin:1999yd}
  D.~Y.~Bardin, P.~Christova, M.~Jack, L.~Kalinovskaya, A.~Olchevski,
   S.~Riemann and T.~Riemann,
  Comput.\ Phys.\ Commun.\  {\bf 133} (2001) 229
  [\hepph{9908433}].


\bibitem{Freitas:2004mn}
  A.~Freitas and K.~M\"onig,
  Eur.\ Phys.\ J.\ C {\bf 40} (2005) 493
  [\hepph{0411304}].


\bibitem{Jersak:1981sp}
  J.~Jersak, E.~Laermann and P.~M.~Zerwas,
  Phys.\ Rev.\ D {\bf 25} (1982) 1218
  [Erratum-ibid.\ D {\bf 36} (1987) 310].


\bibitem{Arbuzov}
 B.~Arbuzov, D.~Y.~Bardin and A.~Leike,
 {\it Mod.\ Phys.\ Lett.} A {\bf 7} (1992) 2029
[Erratum-ibid.\ A {\bf 9} (1994) 1515].

\bibitem{Djouadi}
 A.~Djouadi, B.~Lampe and P.~M.~Zerwas, {\it Z. Phys.} {\bf C67} (1995) 123
 [\hepph{9411386}].


\bibitem{Altarelli}
 G. Altarelli and B. Lampe, \np{B391}{93}{3}.

\bibitem{vanNeerven}
 V. Ravindran and W. L. van Neerven, \pl{B445}{98}{214}.
[\hepph{9809411}].

\bibitem{seymour}
 S. Catani and M.H. Seymour,
 {\it JHEP} {\bf 07} (1999) 023 [\hepph{9905424}].


\bibitem{Bernreuther:1997jn}
  W.~Bernreuther, A.~Brandenburg and P.~Uwer,
  Phys.\ Rev.\ Lett.\  {\bf 79} (1997) 189
  [\hepph{9703305}]; \;
  A.~Brandenburg and P.~Uwer, \np{B515}{98}{279}  [\hepph{9708350}].

\bibitem{Rodrigo:1997gy}
  G.~Rodrigo, A.~Santamaria and M.~S.~Bilenky,
  Phys.\ Rev.\ Lett.\  {\bf 79} (1997) 193
  [\hepph{9703358}]; \;
 \np{B554}{99}{257} [\hepph{9905276}].

\bibitem{Nason:1997tz}
  P.~Nason and C.~Oleari,
  Phys.\ Lett.\ B {\bf 407} (1997) 57 [\hepph{9705295}]; \;
 \np{B521}{98}{237}, [\hepph{9709360}].

\bibitem{Bernreuther:2000zx}
 W.~Bernreuther, A.~Brandenburg and P.~Uwer, \hepph{0008291}.





\bibitem{Gehrmann-DeRidder:2005cm}
  A.~Gehrmann-De Ridder, T.~Gehrmann and E.~W.~N.~Glover,
  JHEP {\bf 0509} (2005) 056
  [\hepph{0505111}].

%


\bibitem{us1}
W.~Bernreuther, R.~Bonciani, T.~Gehrmann, R.~Heinesch, T.~Leineweber, P.~Mastrolia and E.~Remiddi,
Nucl.\ Phys.\ B {\bf 706}, 245 (2005)
[\hepph{0406046}].
%
\bibitem{us2}
W.~Bernreuther, R.~Bonciani, T.~Gehrmann, R.~Heinesch, T.~Leineweber, P.~Mastrolia and E.~Remiddi,
Nucl.\ Phys.\ B {\bf 712} (2005) 229
[\hepph{0412259}].
%
\bibitem{us3}
  W.~Bernreuther, R.~Bonciani, T.~Gehrmann, R.~Heinesch, T.~Leineweber and E.~Remiddi,
Nucl.\ Phys.\ B {\bf 723} (2005) 91
  [\hepph{0504190}].


\bibitem{Catani:1998bh}
  S.~Catani,
  Phys.\ Lett.\ B {\bf 427} (1998) 161
  [\hepph{9802439}].

\bibitem{Sterman:2002qn}
  G.~Sterman and M.~E.~Tejeda-Yeomans,
  Phys.\ Lett.\ B {\bf 552} (2003) 48
  [\hepph{0210130}].

\bibitem{Yennie:1961ad}
  D.~R.~Yennie, S.~C.~Frautschi and H.~Suura,
  Annals Phys.\  {\bf 13} (1961) 379.

\bibitem{Catani:2000ef}
  S.~Catani, S.~Dittmaier and Z.~Trocsanyi,
  Phys.\ Lett.\ B {\bf 500} (2001) 149
  [\hepph{0011222}].

\bibitem{Beneke:1999qg}
  M.~Beneke, A.~Signer and V.~A.~Smirnov,
  Phys.\ Lett.\ B {\bf 454} (1999) 137
  [\hepph{9903260}].
 
\bibitem{Hoang:2001mm}
  A.~H.~Hoang, A.~V.~Manohar, I.~W.~Stewart and T.~Teubner,
  Phys.\ Rev.\ D {\bf 65} (2002) 014014
  [\hepph{0107144}].


\bibitem{Czarnecki:1997vz}
  A.~Czarnecki and K.~Melnikov,
  Phys.\ Rev.\ Lett.\  {\bf 80} (1998) 2531
  [\hepph{9712222}].

\bibitem{Hoang:1997sj}
  A.~H.~Hoang,
  Phys.\ Rev.\ D {\bf 56} (1997) 7276
  [\hepph{9703404}].



\bibitem{AX}
 J.~H.~K\"uhn and T.~Teubner,
  Eur.\ Phys.\ J.\ C {\bf 9} (1999) 221 [\hepph{9903322}].



\bibitem{unknown:2005cc}
The Tevatron Electroweak Working Group, J.F. Arguin et al., 
  ``Combination of CDF and D0 results on the top-quark mass,''
  {\tt hep-ex/0507091}.


%
\bibitem{Kniehl:1989qu}
  B.~A.~Kniehl and J.~H.~K\"uhn,
  Nucl.\ Phys.\ B {\bf 329}, 547 (1990).
%
%

\bibitem{Polylog3}
  T. Gehrmann and E. Remiddi, 
 Comput.\ Phys.\ Commun.\ {\bf 141} (2001) 296
  [\hepph{0107173}]; \;
 D.~Ma\^{i}tre, Comput.\ Phys.\ Commun.\ {\bf 174} (2006) 222 
[{\tt hep-ph/0507152}].





%
\end{thebibliography}
\end{document}